%% file: paper.tex
\documentclass[sigconf,screen]{acmart}

\usepackage[english]{babel}
\usepackage{graphicx}
\usepackage{grffile}
\usepackage{xspace}
\usepackage[normalem]{ulem}
\usepackage{listings}
\usepackage{enumitem}
\usepackage{colortbl}
\usepackage{balance}

\newcommand{\pathfigures}{figures/}

\newcommand{\ie}{i.e.\xspace}
\newcommand{\eg}{e.g.\xspace}
\newcommand{\etal}{\emph{et~al.}\xspace}

\newcommand{\tagger}{\textit{tagger}\xspace}
\newcommand{\silent}{\textit{silent}\xspace}
\newcommand{\ignore}{\textit{forward}\xspace}
\newcommand{\cleaner}{\textit{cleaner}\xspace}
\newcommand{\undecided}{\textit{undecided}\xspace}
\newcommand{\selective}{\textit{selective}\xspace}

\newcommand{\upper}{\textit{upper} field\xspace}

\newcommand{\peer}{\textit{peer}\xspace}
\newcommand{\foreign}{\textit{foreign}\xspace}
\newcommand{\stray}{\textit{stray}\xspace}
\newcommand{\private}{\textit{private}\xspace}

\newcommand{\xti}{\textit{tf}\xspace}
\newcommand{\xtc}{\textit{tc}\xspace}
\newcommand{\xsi}{\textit{sf}\xspace}
\newcommand{\xsc}{\textit{sc}\xspace}
\newcommand{\xtn}{\textit{tn}\xspace}
\newcommand{\xsn}{\textit{sn}\xspace}
\newcommand{\xni}{\textit{nf}\xspace}
\newcommand{\xnc}{\textit{nc}\xspace}
\newcommand{\xnn}{\textit{nn}\xspace}
\newcommand{\xnu}{\textit{nu}\xspace}
\newcommand{\yt}{\textit{t}\xspace}
\newcommand{\ys}{\textit{s}\xspace}
\newcommand{\yi}{\textit{f}\xspace}
\newcommand{\yc}{\textit{c}\xspace}
\newcommand{\yn}{\textit{n}\xspace}
\newcommand{\yu}{\textit{u}\xspace}

\newcommand{\action}{\textit{tagging}\xspace}
\newcommand{\handle}{\textit{forwarding}\xspace}

\newcommand{\downstream}{\textit{downstream}\xspace}
\newcommand{\upstream}{\textit{upstream}\xspace}
\newcommand{\leaf}{\textit{leaf}\xspace}
\newcommand{\transit}{\textit{transit}\xspace}

\newcommand{\condA}{$Cond_1$\xspace}
\newcommand{\condB}{$Cond_2$\xspace}

\newcommand{\cinput}{\textit{input}\xspace}
\newcommand{\coutput}{\textit{output}\xspace}

\newcommand{\pcpair}{$(path,comm)$\xspace}

\newcommand{\ripe}{RIPE\xspace}
\newcommand{\rviews}{RouteViews\xspace}
\newcommand{\pch}{PCH\xspace}
\newcommand{\isolario}{Isolario\xspace}

\newcommand{\datamain}{$d_{May21}$\xspace}

\lstdefinestyle{mystyle}{
    basicstyle=\footnotesize,
    frame=lines,
    breakatwhitespace=false,
    breaklines=false,
    captionpos=b,
    keepspaces=true,
    numbers=left,
    numbersep=10pt,
    showspaces=false,
    showstringspaces=false,
    showtabs=true,
    tabsize=2,
    xleftmargin=5.0ex
}
\copyrightyear{2021}
\acmYear{2021}
\setcopyright{none}
\acmConference[IMC '21]{ACM Internet Measurement
Conference}{November 2--4, 2021}{Virtual Event, USA}
\acmBooktitle{ACM Internet Measurement Conference (IMC '21),
November 2--4, 2021, Virtual Event,
USA}\acmDOI{10.1145/3487552.3487865}
\acmISBN{978-1-4503-9129-0/21/11}

\begin{document}
\title{AS-Level BGP Community Usage Classification}

\title[AS-Level BGP Community Usage Classification]{AS-Level BGP Community Usage Classification}


\author{Thomas Krenc}
\affiliation{%
  \institution{Naval Postgraduate School}
}
\email{tkrenc@nps.edu}
\author{Robert Beverly}
\affiliation{%
  \institution{Naval Postgraduate School}
}
\email{rbeverly@nps.edu}
\author{Georgios Smaragdakis}
\affiliation{%
  \institution{TU Delft}
}
\email{g.smaragdakis@tudelft.nl}

\renewcommand{\shortauthors}{Thomas Krenc et al.}



\input{\pathsections abstract}



\maketitle

\input{\pathsections introduction}
\input{\pathsections related}
\input{\pathsections background}
\input{\pathsections datasets}
\input{\pathsections algorithm}

\input{\pathsections analysis}

\input{\pathsections conclusion}
\input{\pathsections acknowledgements}

\balance
\bibliographystyle{ACM-Reference-Format}
\bibliography{reference}

\input{\pathsections appendix}

\end{document}

%% file: sections/abstract.tex
\begin{abstract}
BGP communities are a popular mechanism used by network operators for
traffic engineering, blackholing, and to realize network policies
and business strategies. In recent years, many research works have contributed to our
understanding of how BGP communities are utilized, as well as how they
can reveal secondary insights into real-world events such as outages
and security attacks.
However, one fundamental question remains unanswered: ``Which ASes
tag announcements with BGP communities and which remove
communities in the announcements they receive?'' 
A grounded understanding of where BGP communities are
added or removed
can
help better model and predict BGP-based
actions in the Internet and characterize the strategies of network
operators.

In this paper we develop, validate, and share data from the first algorithm that 
can infer BGP community
tagging and cleaning behavior at the AS-level.  The algorithm is entirely passive and
uses BGP update messages and snapshots, \eg from public route collectors, as
input.  
First, we quantify
the correctness and accuracy of the algorithm in controlled
experiments with simulated topologies. 
To validate 
in the wild, we announce prefixes with communities
and confirm that more than 90\% of the ASes that we classify behave as
our algorithm predicts.  
Finally,
we apply the algorithm to data from four sets of BGP collectors: RIPE, RouteViews,
Isolario, and PCH. 
Tuned conservatively, our algorithm ascribes
community tagging and cleaning behaviors to more than 13k ASes, the majority
of which are large networks and providers.  We make our algorithm and
inferences available as a public resource to the BGP research
community.
\end{abstract}

%% file: sections/introduction.tex
\section{Introduction}
\label{sec:intro}

Border Gate Protocol (BGP) communities~\cite{rfc1997}, an optional transitive attribute
attached to routing announcements, are widely used to communicate
information and actions within and between Autonomous Systems (ASes).
The flexibility to define communities, tag routes with communities,
and automatically take a prescribed action on aggregates of tagged
routes, has enabled a variety of common uses including traffic
engineering, remote triggered blackholing
(RTBH)~\cite{Cisco-remotely-triggered-black-hole-filtering}, and other custom
network policies and business
strategies~\cite{BenoitOnBGPCommunities}.

Community values are simply byte strings, and there is no globally defined
semantic to these values beyond a handful with special meaning.  ASes
define particular community values either for internal network use,
\eg to tag the geographic location where prefixes are received, or for
signaling to peers, \eg to request AS path prepending.  While it is a
general convention that the high-order bytes of the community are set
to the AS Number (ASN) of the entity defining the community meaning,
the low-order bytes are set according to each network's policy.  And
while some networks publicly publish their community conventions, many
networks do not.  Thus, for both operational and research purposes, it
is often difficult to understand the intent of communities that are
seen attached to routes, for instance at route collectors.

Even more fundamentally, the macro-level community usage behavior
of different ASes is unknown.   As BGP announcements propagate,
individual ASes may arbitrarily ignore, add, modify, or remove
communities.  In this paper, we develop, validate, and employ a
passive algorithm to infer per-AS community usage. At its core, our
algorithm implements a system of constraints that groups ASes
according to those that do or do not tag announcements with
communities, namely, \tagger and \silent ASes, as well those
that remove or ignore communities in updates, namely, \cleaner and
\ignore ASes.  When our algorithm has conflicting information, for
instance due to complex per-peer selective behavior, it
labels the AS as \undecided, and if it cannot make any inference,
\textit{none}.

We first characterize the coverage and accuracy of our algorithm,
along with the sensitivity of its parameters, using simulations of
large AS topologies.  We then validate a subset of our inferences by
using the PEERING testbed~\cite{schlinker2019peering} to announce a prefix under our
control into the BGP.  Using data from three large route collectors,
we show that our algorithm correctly infers the community usage
behavior of 90\% of the ASes we predict, for those ASes in observed
paths.  

Understanding per-AS community usage behavior is important to not
only measurement research that utilizes communities, but also in
uncovering important properties of operational networks in the wild.
For instance, work on measuring community propagation and
security~\cite{IMC2018-Communities-attacks}, community-driven routing load and unnecessary
updates~\cite{krenc2020keep}, community-based outage
detection~\cite{SIGCOMM2017-peering-infrastrutcure-outages}
and blackholing~\cite{IMC2017-blackholing}, and hijack
detection schemes~\cite{sermpezis2018artemis} can all benefit from a grounded per-AS
understanding of BGP community usage behavior.  Thus we see our
work filling an important gap for the measurement and network
research community, akin to AS relationship
inference~\cite{Inferring-Complex-AS-Relationships:IMC2014,caida:as2rel}.
Our contributions include:
\begin{itemize}
 \item A novel passive algorithm to infer per-AS BGP community usage
 behavior.
 \item Analysis of simulated and real-world experiments
 demonstrating the coverage, accuracy, and sensitivity of our 
 algorithm.
 \item Application of our algorithm to data from four major
 route collectors.
 \item We make our algorithm and database of community usage
 behavior publicly available to the network measurement community~\cite{url:communityusage}.
\end{itemize}

The remainder of this paper is organized as follows: In Section~\ref{sec:related}
we introduce related work in this area and provide background information on
BGP and BGP communities in Section~\ref{sec:background}. In
Section~\ref{sec:datasets} we introduce the data sets that we use throughout
this work and the sanitation steps involved. In Section~\ref{sec:inference}, we present our
mental model and derive constraints to implement in to the inference algorithm.
In Section~\ref{sec:inference-validation} we evaluate its performance using
generated ground truth data sets.
In Section~\ref{sec:analysis} we present our classification results using publicly
available routing information. We conclude in Section~\ref{sec:conclusion}.

%% file: sections/related.tex
\section{Related Work}
\label{sec:related}

As the Internet's inter-domain routing glue, the BGP~\cite{rfc4271} is
critical protocol in the operation of the network.  It is therefore
natural that the behavior, evolution, stability, and security of BGP
have been extensively analyzed,
\eg~\cite{dhamdhere2011twelve,griffin1999analysis,10.1145/248156.248160,testart19}
to name a few studies. Similarly, the adoption of BGP
communities~\cite{rfc1997} has driven research into the ways in which
communities are used, as well as how they can reveal important
properties and events in the Internet.

Benoit \etal provided the first taxonomy of BGP
communities~\cite{BenoitOnBGPCommunities}.  Since then, community
usage has continued to increase.  Specifically, among more than 78,000
unique regular communities observed in the recent routing data we analyze,
there are more than 6,300 distinct high-order two-byte values --
indicative of the wide variety of use and breadth of adoption.
Giostas \etal provided one of the first examples of using communities
as a measurement tool, by leveraging observed activity of blackhole
communities to estimate the prevalence, frequency, and duration of
denial-of-service attacks~\cite{IMC2017-blackholing}.  Subsequent work used communities
as a proxy to detect peering infrastructure
outages~\cite{SIGCOMM2017-peering-infrastrutcure-outages}.  

Further motivating our work, Streibelt \etal explored the potential
attack vectors using BGP communities, and their feasibility due to a
general lack of community
filtering~\cite{IMC2018-Communities-attacks}.  Streibelt's work showed
that the lack of common understanding in community values, combined
with their transitive property, leads to communities propagating
further than intended.  As a result, malicious actors in the BGP
system can potentially blackhole or re-route traffic.

Moreover, while Streibelt \etal explore the extent to which communities
propagate, they do not attempt to determine whether individual ASes
are taggers (or silent).  Similarly, Streibelt examines filtering and
propagation of communities on a per-edge basis in the AS graph, but do
not reconcile conflicting information or make per-AS filtering
classifications.  Our algorithm holistically takes into account
filtering and tagging behaviors and is specifically tuned to make
precise inferences.  Last, whereas prior work was entirely
inferential, we use lab experiments, simulations, and live
announcements to validate and understand the limits of our approach.

Also closely related and motivating our investigation, Krenc \etal
explore the BGP update message load and impact due to
communities~\cite{krenc2020keep} and permissive propagation.  In
particular, Krenc's longitudinal study finds a large number of
unnecessary updates that could be avoided, benefiting security and
performance, with stricter community filtering by providers.  

Our algorithm could thus usefully be combined with these prior efforts
to provide a richer understanding of both which ASes are filtering too
permissively, as well as which ASes are vulnerable to attack or
contributing to unnecessary routing traffic.  

%% file: sections/background.tex
\section{Background}
\label{sec:background}

The Internet consists of more than 70k Autonomous Systems (AS).
An AS comprises one to multiple networks under the same administrative
control and it is identified by the AS Number (ASN), a 16-bit integer value.
As with many resource identifiers in the Internet, the ASN is a scarce
resource which limits the number of public ASes in the Internet to $2^{16}$.
Today, while not all 16-bit ASNs are actively routed, the 16-bit address space
is allocated. In order to allow more ASes to participate in the Internet, the
ASN address space has been expanded to 32 bits \cite{rfc6793}.

ASes exchange traffic with each other to allow communication among all Internet users.
Thereby, the traffic flow is determined by contractual business relationships between
pairs of ASes.
In the following, we explain how ASes exchange routing information via the
Border Gateway Protocol (BGP) and introduce the BGP community attribute, which
allows ASes to signal information across the AS level graph.

\subsection{Border Gateway Protocol}
\label{sec:background-bgp}

Business relationships are implemented using the Border Gateway Protocol,
often based on the size and the traffic volumes of the involved ASes.
There exist two primary types of business relationships: peer-to-peer
and provider-customer. ASes typically form peer-to-peer relationships when 
both networks have a similar volume of traffic to exchange.
On the other hand,
in a customer-provider relationship, one of the ASes is the customer who
pays the provider to exchange traffic. This is typically the case, \eg, when a
small network requests transit to the global Internet via a larger network.

Geographically close ASes can interconnect directly at convenient locations and thus
potentially circumvent transit costs and keep traffic local.  Popular peering
locations are Internet Exchange Points (IXP) which provide a shared
infrastructure, large hubs in metropolitan areas, for hundreds of ASes to
exchange traffic directly, either publicly or privately.
In order to facilitate multi-lateral peering, IXPs offer router servers as
value-added service~\cite{IMC2014-RS}. Also, IXPs offer remote peerings where no
physical presence by the AS is required~\cite{giotsas2020peer}. Given the
multitude of possibilities, two ASes can interconnect, it is little surprising
that they can have different business relationships at different locations
which further increases the complexity of
relationships~\cite{Inferring-Complex-AS-Relationships:IMC2014}.

In order for an AS to become globally reachable, it sends prefix announcements
-- encapsulated in BGP update messages -- to neighboring ASes,
which in turn forward the prefix announcements to their neighbors, and so on.
The path a prefix is forwarded along is called the AS path.
An AS path $p$ is a sequence of $n$ ASNs: $A_1, A_2, \dots, A_n$, while
each ASN corresponds to the AS that has forwarded the announcement.

For purposes of exposition in this paper, for any AS path $p$ and AS $A_x \in p$, we
distinguish between \upstream ASes of $A_x$, \ie, all ASes
$A_i \in p,i<x$, and \downstream ASes, \ie, all ASes $A_j \in p, j>x$.
We refer to $A_1$ as the collector \textit{peer}, or simply peer AS,
and $A_n$ as the \textit{origin} AS, while the announcement
originates at $A_n$ and is forwarded \upstream towards $A_1$.

Our inference algorithm relies on views of BGP routes, in particular AS paths
and community attributes, which we obtain via public route collector ``looking
glasses,'' (Section~\ref{sec:datasets}).
Peer ASes forward announcements to route collectors, where the announcements
are archived and made publicly accessible to, \eg, network operators to monitor or debug 
their routing configurations. 

We further distinguish between \leaf and \transit ASes: Given a set of AS paths

$P$, AS $A_x$ is a \leaf AS, $if~\nexists p \in P$ with $A_x \in p \land x \ne |p|$.
Conversely, $A_x$ is a \transit AS, $if~\exists p \in P$ with $A_x \in p \land x < |p|$.
In other words, a leaf AS only originates prefixes but never forwards other prefixes,
because it has no \downstream neighbors. \transit ASes on the other hand forward
announcements from \downstream ASes.

\subsection{BGP communities}
\label{sec:background-communities}

The BGP Communities Attribute~\cite{rfc1997} is appended to updates
and is used to aggregate routes with common properties.
It is a variable-length attribute and can store multiple communities.
Also, the community attribute is a transitive attribute which allows communities to be
propagated across multiple ASes.

A regular BGP community is simply a 32-bit integer that is denoted
in the form $\alpha{:}\beta$, where typically $\alpha$ represents the AS that defines 
the value $\beta$.
Thus, every AS can define its own values without collisions.
Note that this convention applies only to encoding 16-bit ASNs.
In order to accommodate 32-bit ASes as well, the BGP Large Communities Attribute
has been introduced~\cite{rfc8092}.
A large community is a $3 x 32$-bit integer denoted in the form $\alpha{:}\beta{:}\gamma$,
while $\alpha$ represents a 32-bit ASN and $\beta$ and $\gamma$ are additional 64 bits for
the community value.
For simplicity, we refer to $\alpha$ in both community variants
as the \upper in the remainder of this
work\footnote{RFC8092 calls the field the \textit{Global Administrator}.}.
In order to determine the community usage of 32-bit ASes, in this work we
consider large communities as well.
Note that 16-bit ASes can also use large communities.

\textbf{Source of communities:}
Communities can be used to simply convey meta information, \eg, the location
where an announcement has been received. In the following
example, AS $Z$ propagates information to the \upstream AS $X$:

\begin{center}
$ X \xleftarrow{Z{:}*} Z$
\end{center}

$Z{:}*$ is also referred to as \textit{informational} community~\cite{rfc8195}.
However, communities are not always set by the AS that defines it.
For example, communities can instruct other ASes to perform a certain task, \eg, blackholing or path prepending.
In the following example, AS $Z$ instructs AS $X$ to perform a specific action on the announced routes,
indicated by the value defined by AS $X$:

\begin{center}
$ X \xleftarrow{X{:}*} Z$
\end{center}

Here, $X{:}*$ is also referred to as \textit{action} community.

We note that the source of communities can be ambiguous: Some ASes define communities (action and informational) using
a non-public ASN instead of their own in the \upper, e.g., ASes with a 32-bit ASN using
regular (32-bit) communities, or they use standardized, well-defined communities~\cite{rfc1997,rfc8642}.
Further, while route servers at IXPs utilize communities as well, their ASN typically does not appear
in the AS path which further obfuscates the source those communities.

Because the community attribute is an optional attribute, and
BGP provides no mechanisms to validate the source, any AS
along the AS path may add, modify or delete communities.
Thus, it is not guaranteed that the \upper corresponds to the tagging AS.

Our visibility into the BGP system is limited by available collectors
and the ASes to which they peer. 
Since we cannot reliably determine the source of a community, we at a minimum
require its \upper to be present in the associated AS path.
For that purpose, we group communities based on the \upper in relation to the
position of the \upper in the AS path.
We define the following community source groups:
\begin{itemize}
\item A \peer community is a community where the \upper corresponds to the peer
ASN in the AS path, \ie, $A_1$.
\item In a \foreign community, the \upper does not equal the peer's ASN,
but any other ASN in the AS path, \ie, $A_i, i{>}1$.
\item \stray is a community with the \upper representing a public ASN, but the ASN is not in the AS path.
\item \private is a community with the \upper representing a non-public
ASN, \ie, private, reserved, not assigned or allocated, etc.
\end{itemize}
Note, a \peer community in given AS path $p_1$ can appear as a \foreign
community in AS path $p_2$.

Our inference algorithm, which we introduce in
Section~\ref{sec:inference}, necessarily ignores \stray and \private
communities, since there is no indication as to which AS set those
communities without additional knowledge beyond what is available in a
BGP announcement.
For \peer and \foreign communities, we assume the community
was set by the AS that corresponds to the upper 2 bytes (or upper 4 bytes) of the community.
We consider all community types when we verify the correct functioning of
our inference method in Section~\ref{sec:analysis-peerASes}.

\subsection{Community usage}
\label{sec:background-commusage}

We define community usage to describe the way ASes generally configure their
routers to tag routes with communities or filter them,
irrespective of the semantic meaning of community values.

\subsubsection{Mental Model}
In order to better understand community usage in the wild,
we distinguish ASes according to their individual community usage.
To start with, we ask for two basic properties: (1) whether or not an AS
adds its own communities to the community attribute of an announcement,
and (2) whether or not it cleans existing communities received in the 
announcement (set by other
ASes along the AS path).
We refer to the first property as \action behavior and
the second property as \handle behavior.

Since any AS along the AS path can arbitrarily modify the community attribute, we
narrow down the two properties to reflect
consistent behavior, as described in Section~\ref{sec:background-communities}.
For the \action behavior, we define a \tagger AS that defines and sets informational communities internally in a
consistent and automated fashion and forwards them on external BGP sessions.
The counterpart, a \silent AS, does not set its own communities or does not forward them on external sessions.
In addition, we specify the \handle behavior to reflect the forwarding of communities
set by \tagger{s}. Thereby, a \ignore AS forwards communities on external sessions
set by other \tagger{s}.
Conversely, a \cleaner is an AS that does not forward communities on external sessions.
In summary: 

\begin{itemize}
\item \action behavior
\begin{itemize}
\item \tagger: A \tagger AS adds its own communities in a consistent and automated fashion, \ie, informational communities, and forwards them on external links.
\item \silent: A \silent AS may use its own communities internally, but it does not forward them on external links.
\end{itemize}
\item \handle behavior
\begin{itemize}
\item \ignore: A \ignore AS does not remove communities added by other
\tagger ASes and forwards them on external links.
\item \cleaner: A \cleaner AS removes communities set by other \tagger ASes,
either upon receiving or upon forwarding on external links.
\end{itemize}
\end{itemize}

Every AS has a \action and \handle behavior, \eg an AS can be \tagger
and \ignore at the same time (but not \tagger and \silent at the same time).
Note, that our definitions do not exclude the possibility that communities are
used only AS-internally and filtered upon reannouncing.
However, this case is difficult to classify and 
requires AS-specific insights that go
beyond the scope of this work.

\subsubsection{Formal Model}
To capture the \action behavior of AS~$A$ we define a function that returns a
non-empty community set with communities $A{:}*$ if $A$ is a \tagger and an
empty community set if $A$ is a \silent AS:

$\action(A) = 
\begin{cases}
A{:}*,&if~is\_\tagger(A)\\
\emptyset,&if~is\_\silent(A)
\end{cases}
$

Analogously to the \action function, we define a \handle function that captures
the \handle behavior of AS~$A$:

$\handle(A, input) = 
\begin{cases}
input,&if~is\_\ignore(A)\\
\emptyset,&if~is\_\cleaner(A)
\end{cases}
$

Here, the function returns the community set \textit{input} from a neighboring AS if
$A$ is a \ignore, and an empty community set if $A$ is a \cleaner AS.
Thus, the combined community set $\coutput$ of AS $A$ is simply the
union of the \action and \handle functions:

$\coutput(A) = \action(A) \cup \handle(A, \cinput(A))$

When, for example, the \action and \handle functions for AS~$A$ return an empty
set (because $A$ is a \silent and \cleaner AS) then the community set output
$\coutput(A)$ will be empty as well.

In Section~\ref{sec:inference}, we discuss implications of our
mental model and introduce an inference method that
classifies the community usage behavior of an AS~$A$, based on its
community set output $\coutput(A)$.

\subsubsection{Selective tagging}

Thus far, we assume that ASes employ a uniform community usage policy 
for all BGP peers.  The simplest form of community configuration is to 
enable or disable \action and \handle behaviors identically for all neighbors.
Naturally, more complex configurations exist.
A large AS with sessions to multiple ASes may perform different
actions for different neighbors, 
\eg based on the business relationship.
For instance, some ASes
configure their BGP to only propagate communities (\eg geolocation
communities) to customers or to BGP collectors, but not to peers or providers.
Depending on the visibility afforded by different route collectors,
our algorithm can discover such selective community usage behavior.
However, as our primary goal is to definitively classify those ASes
with consistent behavior with high precision, our algorithm labels
these instances as \undecided.

In a worst-case scenario, a peer AS is configured to not forward any communities
to the collectors but to customers or peers. Our algorithm can potentially
misclassify this behavior as \silent, which can have
a detrimental impact on the inferences of other ASes, as we will discuss
in Section~\ref{sec:inference-limitations}.
We explore the performance of our
algorithm in the presence of consistent and selective community behaviors
in Section~\ref{sec:inference-validation}.

%% file: sections/datasets.tex
\section{Data sets}
\label{sec:datasets}

\begin{table*}[t]
  \centering
  \begin{tabular}{|l|r|r|r||r||r|}
\hline
Input data & \ripe & \rviews & \isolario & \datamain & \pch \\\hline
\hline
Entries total & 2,242M & 4,679M & 2,088M & 9,010M & 532M \\
~~~incl. RIB entries & 1,022M & 3,160M & 1,275M & 5,458M & 0\\
Uniq. \pcpair & 46M & 30M & 16M & 77M & 1M\\
\hline
AS numbers & 79,192 & 80,097 & 78,454 & 80,651 & 77,082 \\
After cleaning & 72,737 & 72,837 & 72,625 & 72,951 & 67,541 \\
~~~incl. Leaf ASes & 60,418 & 60,838 & 60,834 & 60,420 & 56,703 \\
~~~incl. 32-bit ASes & 31,147 & 31,193 & 31,082 & 31,239 & 29,077\\
Collector peers & 525 & 291 & 108 & 766 & 1,304 \\\hline
Communities & 12,133M & 16,513M & 11,056M & 39,703M & 2,185M \\
~~~incl. large & 2,614M & 2,196M & 2,283M & 7,093M & 660M \\
Unique communities & 75,681 & 73,720 & 67,816 & 84,186 & 35,804 \\
~~~incl. large & 4,147 & 4,605 & 3,581 & 5,326 & 3,383 \\
Uniq. \upper (regular) & 5,904 & 6,091 & 5,713 & 6,385 & 3,303 \\
Uniq. \upper (large) & 378 & 357 & 358 & 384 & 884\\
Uniq. \upper (both) & 6,160 & 6,340 & 5,958 & 6,643 & 4,292\\
~~w/o \private & 5,660 & 5,739 & 5,474 & 6,025 & 3,931\\
~~~~w/o \stray & 4,316 & 4,373 & 4,189 & 4,579 & 2,574 \\
\hline
  \end{tabular}
  \caption{Data sets overview: RIBs + updates for collector projects \ripe, \rviews, \isolario, \pch in May 19, 2021. \datamain represent the aggregation of \ripe, \rviews, and \isolario. \pch is treated separately due to missing communities in RIB snapshots.}
  \label{tab:datasets-overview}
  \vspace{-5mm}
\end{table*}

Our inferences and analyses utilize AS path /
community set pairs, which we denote as the tuple \pcpair.
We obtain these tuples from public BGP collectors\footnote{BGP collectors used: \textbf{\ripe}: rrc00-26, \textbf{\rviews}: route-views\{2,3,4,6\}, amsix, chicago, chile, eqix, flix, gorex, isc, kixp, jinx, linx, napafrica, nwax, phoix, telxatl, wide, sydney, saopaulo, sg, perth, peru, sfmix, siex, soxrs, mwix, rio, fortaleza, gixa, bdix, bknix, uaeix, \textbf{\isolario}: Alderaan, Dagobah, Korriban, Naboo, Taris, \textbf{\pch}: 236 collectors.}.

Archived BGP update messages contain only a subset of all ASes and AS paths. There exist a significant
number of ASes that do not appear in update data because they do not re-announce their
routes frequently.  These stable ASes and paths can only be found in
RIB snapshots, which are also available from route collectors.
Similarly, updates capture the intermediate changes to the routing system which
are not available from RIBs.
In this work we consider the combined use of archived updates and RIB snapshots.

\subsection{Download and sanitation}
We download RIBs as well as updates encoded in the
Multi-Threaded Routing Toolkit (MRT) format from three major route collector
projects: \ripe~\cite{riperis}, \rviews~\cite{routeviews}, and \isolario~\cite{isolario}
for the complete day May 19, 2021. While we focus mostly on this date, we also
download and analyze other data sets; each time at day granularity. We also obtain \pch~\cite{pch} updates, however,
\pch does not provide RIBs that include the community attribute. To avoid
impairing our inferences due to missing stable ASes from RIBs, we exclude the \pch
data set from most of our analyses.

Before analyzing the routing data, we first filter and transform it
so as not to impart unintentional bias into our results.
We remove routing information that includes unallocated prefixes or ASNs
using current allocation information from the regional registries.
We further remove \textit{AS\_SET}s from AS paths which usually occur when
routes are aggregated.
Also, we prepend the \textit{Peer AS Number} from every MRT message to the
AS path, if the first ASN ($A_1$) does not equal the \textit{Peer AS Number}~\cite{rfc6396}.
This is the case for route servers at IXPs which typically do not participate in
the routing decision process, but still can modify the community attribute.
Finally, we remove path prepending from AS paths by collapsing identical ASNs in succession.

In Section~\ref{sec:analysis} we provide inference results for each individual
collector project. However, for the majority of our analyses, we consider the
aggregated data set of \ripe, \rviews, and \isolario. In the remainder of this
work, we refer to this data set as \datamain.
In Section~\ref{sec:inference} we use the AS paths from the aggregated data set
\datamain in order to generate ground truth data sets to validate
the algorithm.

\subsection{Data overview}
Table~\ref{tab:datasets-overview} provides an overview of the available data sets.
Each of the three data sets -- \ripe, \rviews, and \isolario -- provide us with
Billions of entries 
with around 45\% obtained from RIBs in the case
of \ripe, 67\% \rviews, and 61\% \isolario. 
Note that the route collector projects employ different update file binnings
and snapshot intervals.
There are more than 9G entries in the
aggregate \datamain, of which we extract 77M unique \pcpair pairs.

Overall, the number of distinct ASes observed in \ripe, \rviews, and \isolario are comparable: 
we find approximately 72k ASes which consist of $\sim$60k leaf ASes and $\sim$12k transit ASes.
Furthermore, there exist around 31k
32-bit ASNs in the data set. An AS can be a leaf AS and 32-bit AS at
the same time.
Regarding the number of collector peers, \pch has the highest number
followed by \ripe,
\rviews, and \isolario. Note that a peer AS can peer at multiple BGP collectors.

Lastly, looking at the communities, we observe $\sim$40G communities in
\datamain, of which around 7G stem from the large community attribute.
Overall, there are 84k unique communities.
Those communities exhibit 6,643 unique \upper{s} in both community variations,
while regular communities contribute 6,385 unique \upper{s} and large
communities 384.
Recall from Section~\ref{sec:background-communities}
that 16-bit ASNs can also be encoded in large communities.
Interestingly, \pch shows the highest number of unique \upper{s} in large communities.

As mentioned in Section~\ref{sec:background-communities}, our inference method
has no use for \private and \stray communities.
When we discard communities where the \upper resembles a \private ASN,
6,025 remain.
When we further discard communities where the \upper is never in the
corresponding AS path (\ie \stray), we end up with 4,579 unique \upper{s}.
Those are candidates for \tagger ASes.

%% file: sections/algorithm.tex
\section{Inference Method}
\label{sec:inference}

While some ASes publicly document their community definitions, there is no
documentation as to how those ASes generally set communities on external
sessions, e.g., consistent or selective \action, and whether or not they filter
communities set by other \tagger{s}.
Thus, the community usage has to be inferred.
The purpose of the inference method is to infer the BGP community usage per AS
by observing the community set output.

Recall from Section~\ref{sec:background-commusage} that it is possible to utilize the
community set output $\coutput(A)$ of an AS $A$ to make statements about its
community usage.
Ideally, the more ASes that propagate their community set output to BGP
collectors the better the visibility on the community usage of non-peer ASes.
Unfortunately, only a small fraction of ASes ($\sim$1,751 in \datamain $\cup$ \pch) 
peer with
BGP collectors. The determination of community usage behavior of the remaining
ASes is thus not trivial. Therefore, we devise an inference method that infers
community usage only by utilizing the community set output of BGP collector peers.

In Section~\ref{sec:inference-implications} we state the implications of our
mental model on the inference algorithm, which we detail in Sections~\ref{sec:inference-conditions}
through \ref{sec:inference-algorithm}. Also, we provide a discussion of
our column-based approach in Section~\ref{sec:inference-discussion}.
In Section~\ref{sec:inference-validation} we verify the functioning of the
algorithm by generating and using customized ground truth data sets with known community
usage behavior.

\subsection{Implications from mental model}
\label{sec:inference-implications}

As mentioned above, our inference method is limited to the community set output
of collector peers, \ie,
given a collector $C$, an AS path $p$, and the community set $\coutput(A_1)$ of
the peer AS, such that: $C, A_1, A_2, \dots, A_n | \coutput(A_1)$.
The \action function of any peer AS $A_1$ can be directly observed in its
community set output, because %
$\coutput(A_1) = \action(A_1) \cup \handle(A_1, \cinput(A_1))$.
Consider the following example where ASes $X$ and $Y$ peer with collector $C$:

\begin{center}
$ C \xleftarrow{X{:}*} X $\\
$ C \xleftarrow{\emptyset} Y $
\end{center}

Assuming that every AS uses its own ASN in the \upper of the community, we can
safely conclude that in this context AS $X$ is a \tagger and $Y$ is a \silent
AS.

Further, it is possible to use $\coutput(A_1)$ to infer the
community usage of any AS $A_x{\in}p$.
Note that $\cinput(A_x)$ is the community set $\coutput$ of \downstream neighbor
$A_{x+1}$, \ie: $\cinput(A_x) = \coutput(A_{x+1})$.
Since $\coutput()$ is recursive, $\coutput(A_1)$ can contain information about
the community usage of \downstream ASes.
However, the information can be ambiguous or hidden, as we will explore next.

\subsubsection{Noise}
\label{sec:algorithm-ambiguousinfo}

The simplest form of ambiguity stems from the fact that any AS along the AS path
can arbitrarily modify the communities in the community attribute. The resulting
\textit{noise} can lead to a wrong perception of community usage. Consider the
following example:

\begin{center}
$ C \xleftarrow{Y{:}*} X \xleftarrow{?} Y \xleftarrow{?} Z $
\end{center}

Without additional information beyond what exists in BGP data we are not able
to determine the source of community $Y{:}*$. There are multiple interpretations: (a) AS $Y$ might have tagged the route
with an informational community to indicate the ingress location. (b) AS $Z$
might have used an action community defined by AS $Y$ to instruct $Y$ to perform
a specific action on that route. (c) AS $X$ could have unconventionally
defined and set this community.

If AS $Y$ is a visible \tagger AS (not hidden behind a \cleaner AS), then (b) and
(c) will not impact our classification. If, however, $Y$ is a \silent AS,
(b) and (c) can lead to a misclassification of AS $Y$ to be a \tagger.
In Section~\ref{sec:inference-validation-scenarios} we show that noise, \eg caused by (b) and (c), has no significant
impact on the inference of \tagger and \ignore behaviors, but on the inference of \silent and \cleaner behavior.
To minimize misclassifications, we allow an AS to be labeled \undecided,
if it cannot be clearly assigned \silent or \tagger.

\subsubsection{Hidden behavior}
\label{sec:inference-hidden}

Other forms of ambiguity emerge when the \action and \handle behavior is hidden
behind a \cleaner AS, or when there are no \downstream \tagger ASes to illuminate
the \handle behavior of upstream ASes.
Let us look at an example where AS $X$ has a \downstream \tagger $Z$:

\begin{center}
$ C \xleftarrow{Z{:}*} X \xleftarrow{Z{:}*} Z$
\end{center}

If we see the community $Z{:}*$ at collector $C$, we can conclude that AS $Z$ is a
\tagger while AS $X$ is \silent. More importantly, the fact that we can observe
$Z{:}*$ allows us to conclude that $X$ is a \ignore AS (and not a \cleaner AS).
But it is not always as trivial. Consider now the following example:

\begin{center}
$ C \xleftarrow{\emptyset} X \xleftarrow{?} Z$
\end{center}

Not knowing a priori that $Z$ is a \tagger ($C$ receives an empty community set),
it is impossible for us to tell, whether (a) AS $Z$ is \silent and $X$ is a
\ignore AS, or whether (b) AS $Z$ actually adds communities which are then
removed by \cleaner $X$.

These kinds of ambiguities exacerbate the identification of consistent behavior
as we show next.
Consider the simplified sequence of updates at collector $C$, where $X$ is a \ignore
and $Y$ is a \cleaner AS:

\begin{center}
$ C \xleftarrow{Z{:}*} X \xleftarrow{?} Z $\\
$ C \xleftarrow{\emptyset} Y \xleftarrow{?} Z $\\
$ C \xleftarrow{\emptyset} Y \xleftarrow{?} Z $
\end{center}

From the collector's point of view, one could conclude that $Z$ is a \tagger
because we observe and count the community $Z{:}*$. One could also
count the two occurrences where $Z$ appears silent ($\emptyset$).
Given a configurable threshold, $Z$ can then be classified as either \tagger or \silent.
However, this approach comes with two problems. First, finding the optimal threshold
to work in the wild requires ground-truth which is not available. Second, the approach does not consider the
possibility that $Y$ is a \cleaner AS, erroneously counting those instances as \silent.

Since our goal is to identify consistent behavior with high precision, it is not sufficient to simply count the
occurrences of communities (or the absence thereof).
In addition to the existence of communities, our approach requires $X$ and $Y$
to be a \ignore AS in order to count the behavior of AS $Z$.
Assuming that $X$ is a \ignore AS and $Y$ is a \cleaner in
the above example, we consider only the first update for counting, leading
to the classification of $Z$ as \tagger.

\subsubsection{AS-level periphery}
\label{sec:inference-periphery}
Another form of hidden behavior is shown by leaf ASes (as defined in
Section~\ref{sec:background-bgp}). Leaf ASes have
no \downstream ASes and thus no $\cinput()$ to determine their \handle behavior.
We consider leaf ASes useful since their \action behavior can still help
to illuminate the \handle behavior of \upstream ASes.

\subsection{Conditions}
\label{sec:inference-conditions}

In order to deal with ambiguous or hidden information, and subsequently minimize wrong
inferences, in the following we define conditions that fit our mental model and are required
when counting the community usage per AS, see Section~\ref{sec:inference-counting}.
If the conditions are not true, we cannot make any statement about the community
usage of an AS $A_x$, given a \pcpair pair $C, A_1, A_2, \dots, A_n | \coutput(A_1)$.

\begin{center}
\textbf{Cond1:} $is\_\ignore(A_i)|\forall{A_i}, i{<}x$
\end{center}

In order to make any statement about AS $A_x \in path$ we need to make sure that
all \upstream ASes $A_i$, with $i{<}x$ are \ignore. %
Conversely, when a single \upstream AS is a \cleaner, \eg, if $A_{x-1}$ is a \cleaner AS, then %
$\coutput(A_{x})$ is hidden. %
Thus, $Cond_1$ must be fulfilled in order to determine the \handle and
\action behavior of AS $A_x$ ($\handle(A_x,input)$ and $\action(A_x)$).

\begin{center}
\textbf{Cond2:} $is\_\ignore(A_j){\land}is\_\tagger(A_t)|\forall{A_j}\exists{A_t},x{<}j{<}t$
\end{center}

In order to determine the \handle behavior of AS $A_x$, %
an additional condition is required. Recall from Section~\ref{sec:algorithm-ambiguousinfo}
that in order to determine the \handle behavior of an AS, it requires input from a
\downstream \tagger, here $A_t$.
Thus, if there are no \downstream \tagger{s}, or the \downstream \tagger{s} are hidden
by a cleaner, we cannot make any statement about the \handle behavior of $A_x$.
For example, assume $is\_\silent(A_{x+1}) \land is\_\cleaner(A_{x+1})$ is true,
then $\coutput(A_{x+1}) = \emptyset$ and $\handle(A_{x}, \emptyset) = \emptyset$.

\subsubsection{Race conditions}
\label{sec:inference-racecond}
Under certain topological conditions, both the \action and \handle behavior
of ASes cannot be determined. Consider the following scenario: AS $X$ is the only \upstream
provider of AS $Y$, while $Y$ is the only \downstream AS of $X$.

\begin{center}
$ C \xleftarrow{?} X \xleftarrow{?} Y$
\end{center}

In order to infer the \handle behavior of AS $X$, at least one \downstream AS
is required, here AS $Y$, to be a visible \tagger (\condB). However, to
count the \action behavior of AS $Y$ in the first place, it requires all \upstream ASes,
here AS $X$, to be \ignore (\condA). Thus, AS $X$ requires pre-existing
knowledge about AS $Y$ and vice versa which our inference method cannot resolve.

\subsubsection{Summary}

The community set $\coutput(A_1)$ allows us to determine the \action and \handle behavior
of ASes $A_i \in $ path $p$. However, noise, as well as \silent and \cleaner
ASes that hide information about other ASes, can lead to ambiguities.
In order to minimize wrong inferences, the
\action function requires $Cond_1$, and the \handle function requires both,
$Cond_1$ and $Cond_2$ to be true.
However, race conditions can occur which leave some ASes without class, \ie, \textit{none}.
In Section~\ref{sec:inference-validation-scenarios} we measure the impact of
race conditions on our inference coverage.

\subsection{Counting community usage}
\label{sec:inference-counting}

In order to infer the community usage per AS,
we observe and count patterns that satisfy requirements derived from the implications in Section~\ref{sec:inference-implications}.
Given input $C, A_1, A_2, \dots, A_n | \coutput(A_1)$,
we increase the class counter for a given AS $A_x$ if the following conditions hold true:

\begin{itemize}
\item \action
\begin{itemize}
\item $\yt[A_x]\mathbin{++}~if~Cond_1~\land~{A_x}{:}* \in \coutput(A_1)$
\item $\ys[A_x]\mathbin{++}~if~Cond_1~\land~{A_x}{:}* \notin \coutput(A_1)$
\end{itemize}
\item \handle
\begin{itemize}
\item $\yi[A_x]\mathbin{++}~if~Cond_1~\land~Cond_2~\land~{A_t}{:}* \in \coutput(A_1)$
\item $\yc[A_x]\mathbin{++}~if~Cond_1~\land~Cond_2~\land~{A_t}{:}* \notin \coutput(A_1)$
\end{itemize}
\end{itemize}

Note that $Cond_1$ drops out if $x{=}1$, i.e., it is trivial to determine
the \action function of peer ASes. Also, if conditions for \handle counters
$\yi[]$ and $\yc[]$ are met, the conditions for \action counters are met as well.
If $Cond_1$ or $Cond_2$ are not met, none of the counters is increased.
Based on those counters, we can query the \action and \handle behavior of $A$:

\begin{itemize}
\item \action
\begin{itemize}
\item $is\_\tagger(A) = \frac{\yt[A]}{\yt[A] + \ys[A]} \geq \tagger\_thrsh$
\item $is\_\silent(A) = \frac{\ys[A]}{\yt[A] + \ys[A]} \geq \silent\_thrsh$
\end{itemize}
\item \handle
\begin{itemize}
\item $is\_\ignore(A) = \frac{\yi[A]}{\yi[A] + \yc[A]} \geq \ignore\_thrsh$
\item $is\_\cleaner(A) = \frac{\yc[A]}{\yi[A] + \yc[A]} \geq \cleaner\_thrsh$
\end{itemize}
\end{itemize}

For example, an AS A is a \tagger if the share of $\yt[A]$ exceeds the \tagger
threshold. Recall from Section~\ref{sec:background-commusage} that a \tagger is an AS that
sets its own communities in a consistent manner. Thus, we want the threshold to be
as high as possible, but at the same time allow for exceptions, \eg, 99\%.
We explore different thresholds in Section~\ref{sec:inference-validation}.

\subsection{Selective behavior}
\label{sec:inference-limitations}

\label{sec:inference-limitations-selective}

Not all ASes have a uniform \action and \handle behavior.
An AS may add own and remove other communities selectively, e.g., on a
per-session basis or based on the business relationship.
Selective \action can lead to a contradicting
perception of community usage, \ie, the \tagger counter $\yt[Z]$ for a given AS
$Z$ is increased in one instance, the \silent counter $\ys[Z]$ for the same AS may be increased
in another. Consider the following two updates, where $X$ and $Y$ are \ignore ASes:

\begin{center}
$ C \xleftarrow{Z{:}*} X \xleftarrow{Z{:}*} Z $\\
$ C \xleftarrow{\emptyset} Y \xleftarrow{\emptyset} Z $
\end{center}

We assume that AS $Z$ is a
\selective \tagger that adds communities only to customers, here AS $X$, but
not to peers (AS $Y$).
Based on that observation we increase the \tagger and \silent counters for the same AS $Z$.
When both \action counters $\yt[Z]$ and $\ys[Z]$ are such
that neither $is\_\tagger(Z)$ nor $is\_\silent(Z)$ is true, we simply
classify $Z$ as \undecided, see Section~\ref{sec:inference-classification}.

While \undecided ASes invalidate \condA and \condB and thus exacerbate further
inferences, misclassifications due to selective tagging can lead to further
misclassification, as we show next.
Consider the following
scenario, where AS $Z$ always tags routes toward the collector using its own
communities, but never to any other AS:

\begin{center}
$ C \xleftarrow{Z{:}*} Z $\\
$ C \xleftarrow{\emptyset} X \xleftarrow{\emptyset} Z $
\end{center}

If there are sufficient occurrences of the first update in the input data, AS $Z$ will be
classified as \tagger. Now, assume that the algorithm attempts to
determine the \handle behavior of $X$ using the second update. If there are
sufficient occurrences visible, the algorithm will classify
$X$ as \cleaner, because $is\_\tagger(Z)$ is true, and the community $Z{:}*$ is not present.

\subsubsection{Conclusion}
Selective community usage is a root cause that
leads to a limited view of community usage behavior. They can
invalidate conditions \condA and \condB and thus exacerbate the counting.
Subsequently, the missing counts of one AS can have a detrimental impact on the
counting of another AS, etc.
In Section~\ref{sec:inference-validation-scenarios} we show how our algorithm
deals with selective \action.

\subsection{Classification}
\label{sec:inference-classification}

After the counting is done, we determine the community usage behavior, i.e., the
\action and \handle behavior of AS~$A$ like this:

$get\_class(A) = get\_\action(A) || get\_\handle(A)$

$get\_\action(A) = 
\begin{cases}
\yn,&if~\yt[A] + \ys[A] == 0\\
&else
\begin{cases}
\yt,&if~is\_\tagger(A)\\
\ys,&elif~is\_\silent(A)\\
\yu,&else
\end{cases}
\end{cases}
$

$get\_\handle(A) = 
\begin{cases}
\yn,&if~\yi[A] + \yc[A] == 0\\
&else
\begin{cases}
\yi,&if~is\_\ignore(A)\\
\yc,&elif~is\_\cleaner(A)\\
\yu,&else
\end{cases}
\end{cases}
$

The function $get\_class(A)$ returns a two-character string which indicates
the inferred \action (first character) and \handle (second character) behavior
of AS~$A$.

Both indicators can assume \yn (none) if the respective counters are zero.
Recall, when any of the required conditions $Cond_1$ or $Cond_2$ cannot be fulfilled,
none of the counters are increased. Thus, it can occur that all counters for an
AS result in zero.
When both \action counters $\yt[A]$ and $\ys[A]$ for a given AS $A$ are zero,
there is not enough information to make any decision about the \action
behavior of $A$. The same applies to \handle counters.

When \action counters are available, the resulting \action indicator can be
\yt (\tagger), \ys (\silent), or \yu (\undecided).
Similarly, the \handle indicator can be either \yi (\ignore),
\yc (\cleaner), or \yu. Note, that the resulting class depends on
the thresholds that are used during counting, see Section~\ref{sec:inference-counting}.

The resulting two-character string for AS~$A$ can range among
\xti (\tagger-\ignore),
\xsc (\silent-\cleaner),
or \xnu (\textit{none}-\undecided).

\subsection{Algorithm}
\label{sec:inference-algorithm}

\begin{figure}[t]
\centering
\includegraphics[width=.70\linewidth]{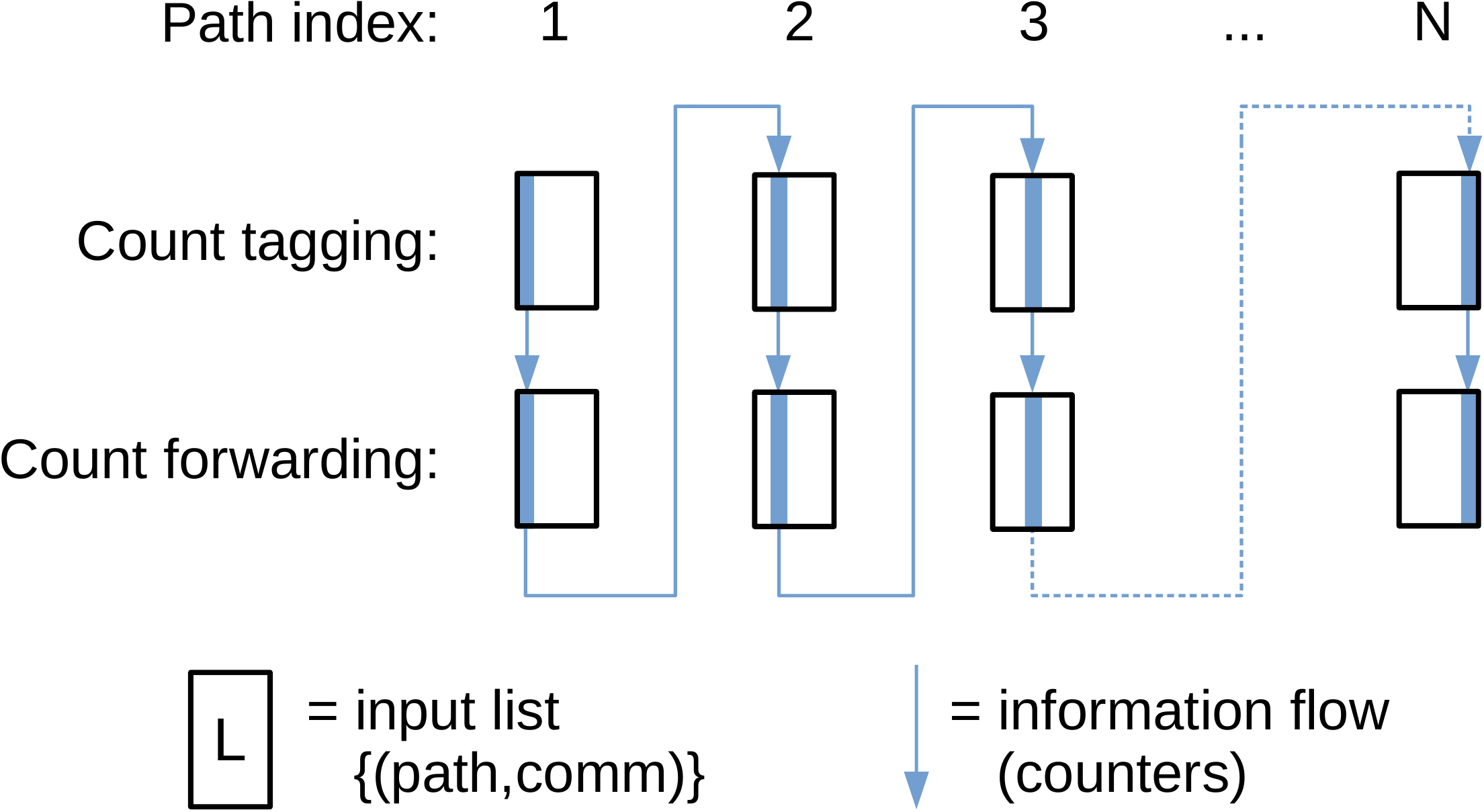}
\caption{Inference algorithm workflow. Two passes over input \pcpair tuples: 1) to count \action behavior and 2) to count \handle behavior for each AS. Tuples are iterated per column (AS path index), starting with left-most ASes ($A_1$). Knowledge from one iteration is transferred to the next.}
\label{fig:algo_main}
\end{figure}

Considering the requirements for counting community usage (see Section~\ref{sec:inference-counting})
we note that counting each class depends on pre-existing knowledge of other ASes
in the AS path, i.e., $Cond_1$ and $Cond_2$.
In particular counters $\yc[]$ and $\yi[]$ require for conditions to hold true
for \upstream ASes as well as \downstream ASes.
Fortunately, it is trivial to count the \action behavior of peer ASes ($A_1$),
which can be used as initial knowledge to feed follow-up inferences.
In order to meet the conditions $Cond_1$ and $Cond_2$ as often as possible,
we design an algorithm that generates pre-existing knowledge by iterating over
the input data a total of two times, in a column-based fashion (as opposed to row-based), see Figure~\ref{fig:algo_main}.
Iterating the data by column, \ie, the path index, we can utilize the recursive output of $A_1$ and
apply the conditions as the algorithm progresses to higher indices.
To show a sample implementation of the algorithm, we provide the corresponding pseudo-code in Listing~\ref{lst:algorithm} in the appendix.

In the following, we describe the steps of the algorithm in detail.
Given as input $L$ = \{$\pcpair$\}, a list of AS path and
community set tuples, where $path = A_1,A_2,\dots,A_n$ and
$comm = \coutput(A_1)$, we begin with counting the \action behavior of peer ASes:

\begin{itemize}
\item path index $1$
\begin{itemize}
\item \textbf{count \action} We begin by counting the \action behavior at path index $1$,
\ie, $A_1$ of every AS path.
For that,
we increase the \tagger counter $t[A_1]~if~{A_1}{:}* \in \coutput(A_1)$.
Otherwise, we increase the \silent counter $s[A_1]$.
Note that \condA drops out at path index $1$, since there are no upstream
ASes.

\item \textbf{count \handle}
Next, we use the previously obtained \action counters to fulfill $Cond_2$ of
\handle counters $i[]$ and $c[]$, and count the \handle behavior at path index $1$.
We increase the \ignore counter $i[A_1]~if~{A_t}{:}* \in \coutput(A_1)$. Else,
we increase the \cleaner counter $c[A_1]$.
\end{itemize}

\item path index $i>1$ ($Cond_1$ kicks in)
\begin{itemize}
\item \textbf{count \action}
We use \handle counters from previous steps to fulfill $Cond_1$ and count
the \action behavior at path index $i$.
We increase $t[A_i], if {A_i}{:}* \in \coutput(A_1)$. Otherwise, we increase $s[A_i]$.

\item \textbf{count \handle}
We use \handle and \action counters from previous steps to fulfill $Cond_1$ and
$Cond_2$ and count the \handle behavior at path index $i$.
We increase $i[A_i], if {A_t}{:}* \in \coutput(A_1)$. Otherwise, we increase $c[A_i]$.
\end{itemize}

\end{itemize}

Summary: In essence, the inference method iterates over the input list $L$ two
times, \ie, $2 * (N*|L|)$, where $N$ corresponds to the maximum path index.
For each path index (column) we first count the \action behavior, then the
\handle behavior.
With each iteration over a path index, the gained knowledge is transferred to the
next iteration.
We note that at high path indices the number of ASes that need to fulfill
$cond_1$ increases, and thus, the probability of inference decreases.
Although the maximum observed path length in the input data is 19 (after cleaning),
we observe that the algorithm stops increasing counters, \eg, due to missing
information, typically at around $N{=}7$.

\subsection{Discussion}
\label{sec:inference-discussion}

In order to apply rules that require pre-existing knowledge and at the same
time traverse the input data efficiently, we chose an approach that considers
both.
Instead of traversing the entire paths one by one (row-based), in our approach we
traverse all paths by the path index (column-based), starting with the collector
peer.
 Recall, $\coutput(A_1)$ is recursive and thus
can contain the community output of other ASes along the path.

\textbf{Row-based}: In a row-based approach, each path is processed separately and independently
from the other paths. Thus, the counters for observing communities (or the lack thereof)
associated with an AS path cannot be connected to observations in other paths.
Inferences can be made based on the counters only after all paths are processed.
In order to achieve our objectives, \ie, classifying consistent behavior with
high precision, additional refinement and optimization steps would be required.

Further, since the row-based approach counts observations without pre-existing
knowledge about the ASes in the AS path, it is susceptible to misclassification,
\eg due to noise or hidden behavior. Also, since there is no ground truth
information available about the community usage in the wild, finding the right
threshold to adjust the precision is not possible. In the appendix, we provide
pseudo-code that serves as an example for a row-based approach, see Listing~\ref{lst:algorithm-alter}.

\textbf{Column-based:} Like the row-based approach, our column-based approach allows us to count while we traverse the 
AS paths. However, in addition, it can generate pre-existing knowledge
at one path index and used it at later indices.
This further enables us to implement rules, \ie \condA and \condB that
base on our mental model, on every AS path.
Thus, our algorithm is able to identify

\begin{itemize}
\item ASes which behavior is hidden behind a \cleaner AS and
\item ASes that are not illuminated by a \downstream \tagger.
\end{itemize}

Both abilities allow us not only to minimize misclassification, but also
they increase the robustness of inferences in the presence of noise.
In Section~\ref{sec:inference-validation} we show that our
algorithm avoids misclassifications and that it classifies less than 0.5\% of
ASes that are actually hidden.

\section{Verification of algorithm}
\label{sec:inference-validation}

In this section, we apply the algorithm to different scenarios in controlled
simulations, in order to verify that it performs according to our expectations.
We generate multiple data sets with known community usage behavior. Therefore,
we take all available AS paths from \ripe, \rviews, and \isolario (\datamain),
assign roles to each AS, e.g., \xti (\tagger-\ignore), \xsc (\silent-\cleaner),
and compute the community output of each peer AS ($\coutput(A_1)$), according to
our mental model in Section~\ref{sec:background-commusage}. Thus, we have an AS
path substrate from real-world BGP data augmented with known community usage
behavior per AS. Specifically, we know which AS's behavior is consistent
(\tagger, \ignore, ...) and selective and which behavior is always hidden and
thus cannot be inferred.
Further, each scenario uses the AS paths from \datamain as a substrate. Thus, each
generated data set includes 72,951 ASes among which 60,420 are leaf ASes and 766 are collector peers. All
presented results were inferred using a threshold of 99\% in order to increase
the inference counters.

\subsection{Consistent behavior}
\label{sec:inference-validation-consistent}

To test our algorithm against the average and extreme cases, we generate ground
truth data sets with different settings in the community usage behavior of ASes.
Initially, we consider consistent behavior of ASes, e.g., a \tagger adds
communities towards all BGP neighbors (including collectors), irrespective of the
business relationship.

\begin{itemize}
\item \textbf{alltf:} To test how the algorithm performs under optimal
conditions, in this scenario, we maximize the visibility of communities by
assigning the role \xti to all ASes. This way the \action and \handle behavior
of all ASes becomes visible.
\item \textbf{alltc:} In this scenario, we test the opposite case, i.e., where
the visibility of communities is minimized. Therefore, the role \xtc is assigned
to all ASes.
\item \textbf{random:} Since there exist no ground truth information regarding the community
usage of ASes, we create an average case scenario, where all roles \xti, \xtc,
\xsi, and \xsc are uniformly assigned at random among all ASes, \ie, 25\% are \xti,
25\% \xtc, 25\% \xsi, and 25\% \xsc.
\end{itemize}

Intuitively, the algorithm should perform best in the \textit{alltf} scenario
and perform worst in the \textit{alltc} scenario. Note, since each AS behaves
consistently, the \action and \handle behavior, if not hidden, will be inferred
correctly, irrespective of the threshold.

In order to better understand the sensitivity of our inference method,
we add additional communities to the generated ground truth
data sets that stress the ability of the algorithm to infer the \action and
\handle behavior of ASes correctly.

\begin{itemize}
\item \textbf{noise:} First, to test the \action behavior inference, we let
around 50\% of all ASes set communities using the ASN of their \upstream neighbor in the
\upper, simulating action communities. Second, to test the \handle inference,
we add a community with the \upper corresponding to the ASN of the originator.
\end{itemize}

These two noise sources occur with a 5\% probability, such that each affected AS
can exhibit inconsistent behavior.
In the next section, we introduce
additional scenarios in which we explore the \selective behavior of ASes.

\subsection{Selective behavior}

Since not all ASes exhibit a consistent community usage behavior, their \action
and \handle counters may lead to an \undecided inference, or even to a
misclassification, see Section~\ref{sec:inference-limitations-selective}.
In order to test how the algorithm reacts to \selective behavior,
we use the \textit{random} scenario with consistent behavior and modify around 50\% of
the assigned \tagger ASes to selectively tag routes based on the business
relationship:

\begin{itemize}
\item \textbf{random-p:} In this scenario, the \selective \tagger
ASes do not set communities on provider links, but on all other links, \ie,
peers, customers and collectors.
\item \textbf{random-pp:} In this scenario, the \selective \tagger ASes
do not set communities on provider and peer links; only towards and customers
and collectors.
\end{itemize}

For our convenience, we use CAIDA's business relationship inferences
\cite{caida:as2rel} to augment the ground truth data sets with the corresponding
behavior. For example, a \selective \tagger AS $A_x$ in scenario \textit{random-pp} adds a
community towards its neighbor $A_{x-1}$, only if $A_{x-1}$ is a customer of
$A_x$. Since we create a hypothetical scenario, we do not rely on the accuracy
of the business relationship inferences.

Other than that, the parameters of the \textit{random} scenario of Section~\ref{sec:inference-validation-consistent}
are used. Since in \textit{random-p} and \textit{random-pp} the propagation of
communities is limited, the algorithm is expected to perform worse compared to
the \textit{random} scenario.

\subsection{Results}
\label{sec:inference-results}

\begin{table*}[t]
\small
  \centering
  \begin{tabular}{|l||r|r|r|r||r|r|r|r||r|r|r|r||r|r|r|r|}
\hline
 & \multicolumn{2}{c|}{\action} & \multicolumn{2}{c||}{\handle} & \multicolumn{4}{c||}{full classification} & \multicolumn{4}{c||}{partial classification} & \multicolumn{4}{c|}{none / undecided}\\\hline
scenario  & rec. & prec. & rec. & prec. & \xtc & \xsc & \xti & \xsi & \xtn & \xsn & \xnc & \xni & \xnn & u* & *u & uu\\\hline
\multicolumn{17}{l}{consistent behavior:}\\\hline
alltc & 1.00 & 1.00 & 0.82 & 1.00 & 578 & 0 & 0 & 0 & 188 & 0 & 0 & 0 & 72,185 & 0 & 0 & 0\\
alltf & 0.96 & 1.00 & 0.83 & 1.00 & 0 & 0 & 10,427 & 0 & 59,570 & 0 & 0 & 0 & 2,954 & 0 & 0 & 0\\
random & 0.93 & 1.00 & 0.70 & 1.00 & 1,295 & 1,298 & 1,319 & 1,295 & 20,558 & 20,613 & 0 & 0 & 26,573 & 0 & 0 & 0\\
random+noise & 0.55 & 1.00 & 0.45 & 1.00 & 459 & 242 & 1,256 & 618 & 20,067 & 3,282 & 1 & 1 & 27,808 & 17,518 & 1,288 & 412 \\\hline
\multicolumn{17}{l}{\selective behavior:}\\\hline
random-p  & 0.42 & 0.86 & 0.39 & 0.97 & 623 & 766 & 403 & 482 & 7,275 & 13,513 & 1 & 1 & 48,980 & 622 & 270 & 16\\
random-pp  & 0.18 & 0.89 & 0.18 & 0.94 & 329 & 399 & 134 & 169 & 3,484 & 5,817 & 0 & 1 & 62,035 & 286 & 288 & 12\\
\hline
  \end{tabular}
  \caption{Classification results and performance using scenarios with consistent and \selective behavior. The numbers represent mean values from 10 iterations per \textit{random} scenario. Thresholds 99\%, ground truth data substrate: \datamain, 72,951 ASes (60,420 leafs, 31,239 32-bit, 766 collector peers), in May 19, 2021.}
  \label{tab:classresults-groundtruth}
\end{table*}

In Table~\ref{tab:classresults-groundtruth} we present the classification results
for the scenarios with consistent behavior (\textit{random}, \textit{alltf},
\textit{alltc}), and those with \selective behavior (\textit{random-p},
\textit{random-pp}).
In each \textit{random} scenario, we use 10 different ground truth data sets, each
generated with a different (random) distribution of roles (\xti, \xtc, \xsi, \xsc).
Since, the resulting inferences are comparable in all 10 variations, we only show the
mean values in the table.

For simplicity, we calculate the recall by considering only \action
and \handle behaviors that are not selective, hidden or missing (\ie, \handle
behavior at leaf ASes). Thereby, the false negatives include cases where
the \action or \handle behavior was not inferred, \ie, \textit{none}, or the
inference results in \undecided.

Let us first look at scenarios with consistent behavior:
\begin{itemize}
\item \textbf{Precision and recall:} All scenarios with consistent behavior (including random+noise) show a precision of 100\%, \ie,
there is no wrong inference of \action and \handle behavior. Also, the recall
is relatively high when inferring \action (93-100\%) and \handle (70-82\%).
However, the recall in the case of random+noise is low, \ie, 55\% for \action and 45\% for \handle behavior.
\item \textbf{\textit{full classification}:} In the \textit{random} scenario, the algorithm yields around 1,300 inferred ASes
for each of the assigned roles, \ie, both, the \action and \handle
behaviors were inferred. Further, in the \textit{alltc} scenario 578 ASes are
\xtc and in the \textit{alltf} scenario 10,427 ASes are \xti.
\item \textbf{\textit{partial classification}:}
A significant amount of ASes have an inferred \action behavior but no \handle behavior.
$\sim$20k ASes are identified as \tagger and
$\sim$20k as \silent in the \textit{random} scenario.
\item \textbf{\textit{none / \undecided}:} Looking at the number of not classified ASes, the
algorithm performs best in the \textit{alltf} scenario, and worst in the
\textit{alltc} scenario, as expected. Notably, the random+noise scenario leads
to \undecided behavior in particular the \action behavior is affected, see u* (\action is \undecided).
\end{itemize}

Next, we look at the \selective behavior of ASes, \ie, {\textit{random-p}} and
\textit{random-pp}.
Compared to the simple \textit{random} scenario, the \selective behavior of ASes leads to an
increase of \undecided inferences in the case of \action and \handle,
see u* (\action is \undecided) and *u (\handle is \undecided) in Table~\ref{tab:classresults-groundtruth}.
Note that \undecided behavior exacerbates the inference of other ASes, see Section~\ref{sec:inference-limitations}.
Unsurprisingly, the distribution of inferences is skewed compared to the simple \textit{random} scenario.
Also, when we consider the \textit{none} cases as a performance metric, we find
that the algorithm still performs better than the \textit{alltc} scenario and
worse than the \textit{alltf} scenario.

The most important difference resulting from those scenarios is recall and precision.
Because of \selective \tagger{s} the precision of our inference method in both scenarios is affected,
\ie, 86\%/99\% in the case of \action and 97\%/94\% in the case of \handle. The
recall is more affected: 42\%/18\% in the case of \action and 39\%/18\% in the
case of \handle. Given the increased difficulty in the \textit{random-p} and
\textit{random-pp}, these results are expected.

\subsubsection{Performance under different thresholds}

Selective \action behavior  can lead to a contradicting
perception of community usage, \ie, the \tagger counter $\yt[Z]$ for a given AS
$Z$ is increased in one instance, the \silent counter $\ys[Z]$ for the same AS may be increased
in another.
In the following, we show how the sensitivity and
specificity of our inferences change when different thresholds are used to infer
the \action or \handle behavior of an AS. We
repeat the inference method on both scenarios, \textit{random-p} and \textit{random-pp},
for every threshold between 50\% and 100\%.

\begin{figure}[t]
\centering
\includegraphics[width=.49\linewidth]{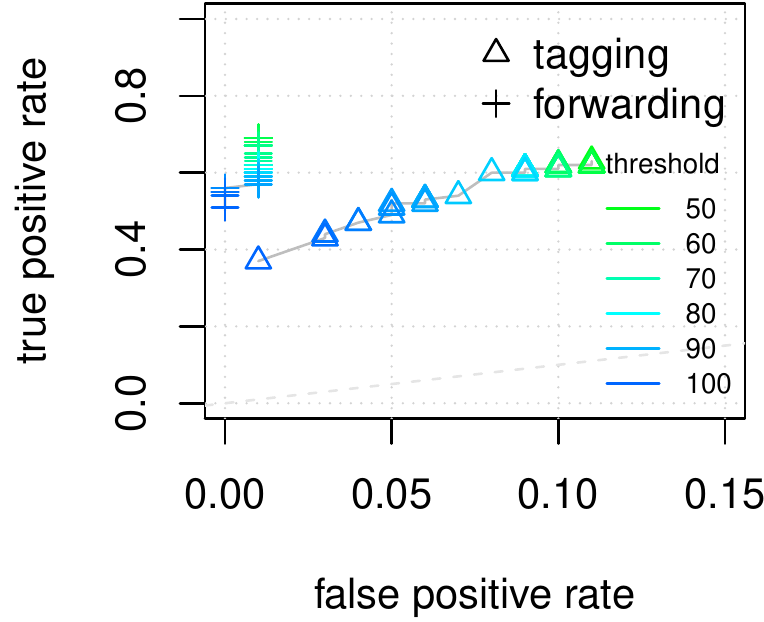}
\includegraphics[width=.49\linewidth]{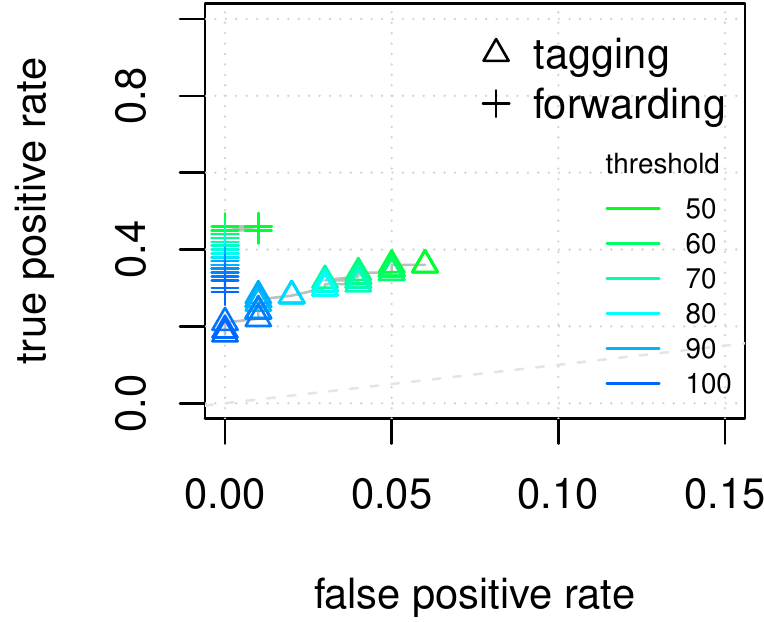}
\caption{ROC curves show effects when changing threshold. In scenario \textit{random-p} (left) \tagger{s} add communities on customer and peering links, and in scenario \textit{random-pp} (right) \tagger{s} add communities on customer links only (increased difficulty). Performance not sensitive to threshold.}
\label{fig:inference-randomp-roc}
\end{figure}

Figure~\ref{fig:inference-randomp-roc} show
the ROC curves for the \action and the \handle inferences, for the scenarios
\textit{random-p} (left) and \textit{random-pp} (right), respectively.
The false-positive rate is on the x-axis and the true-positive rate on the y-axis.
As we can see, by increasing the threshold from 50\% to 100\% the specificity
increases and the sensitivity decreases. However, those are minute changes:
The false-positive rate decreases from 10\% to 1\% in case of the \action
classifier, and from 1\% to 0\% in case of the \handle classifier.
These results indicate that the inferences are not very sensitive to the
threshold, in particular when inferring the \handle behavior.
Also, the true-positive rate decreases by $\sim$20\% in both cases.
Since the difficulty in scenario \textit{random-pp} is increased, the number of
undecided ASes increase as well. Thus, we observe a lower true-positive rate
compared to \textit{random-p}.

\subsection{Scenario Impact}
\label{sec:inference-validation-scenarios}

In this section, we explain the impact of the different
scenarios on the inference of \action and \handle behavior, respectively.
Thereby, we contrast the assigned roles against classification results
with the help of confusion matrices, as provided in the tables below.
Where necessary, we also include the information
\textit{(hidden)} if the assigned role is hidden (Section \ref{sec:inference-hidden}),
and \textit{(leaf)} if the \handle behavior is assigned to a leaf AS (Section~\ref{sec:inference-periphery}).
Recall that each scenario uses the AS paths from \datamain as a substrate which
includes 72,951 ASes, among which are 60,420 leaf ASes and 766 collector peers.
In the following, we provide a selection of results. For a complete list,
see Tables~\ref{tab:confusion-action} and~\ref{tab:confusion-handle}
in the appendix.

\textbf{alltf}: Let us consider the scenario \textit{alltf} where all ASes are
assigned the role \tagger-\ignore and the visibility is maximized:

\setlength{\intextsep}{2mm}

\begin{table}[H]
\small
  \centering
  \begin{tabular}{|l||r|r|r|r|}
\multicolumn{1}{l}{assigned roles:} & \multicolumn{4}{c}{classification result:}\\\hline
alltf:		& \tagger & \silent & \undecided &   none\\\hline
\tagger		& 69,997 &      0 &         0 &  2,954\\\hline
  \end{tabular}
\end{table}

The algorithm is able to attribute a correct \action behavior to more than 95\% of the ASes.
For the remaining 2,954 ASes, the algorithm cannot generate pre-existing
knowledge for neighboring ASes due to the race condition
limitation of our algorithm as described in Section~\ref{sec:inference-racecond}.
Thus, there are no counters and the inference returns \textit{none}.

When we look at the \handle behavior in the same scenario, we see that more than
60k ASes are leaf ASes, and are therefore not attributed a \handle behavior:

\begin{table}[H]
\small
  \centering
  \begin{tabular}{|l||r|r|r|r|}
\multicolumn{1}{l}{assigned roles:} & \multicolumn{4}{c}{classification result:}\\\hline
alltf: 	   & \ignore & \cleaner & \undecided &   none\\\hline
\ignore		   &  10,427 &        0 &          0 &  2,104\\
\ignore (leaf)	   &       0 &        0 &          0 & 60,420\\\hline
  \end{tabular}
\end{table}

 Given those
ASes that are not leaf, the algorithm is able to correctly classify more than
83\% (10,427) as \ignore.

\textbf{alltc}: In this scenario, all ASes are \tagger-\cleaner.
Since all ASes
are \cleaner{s}, we correctly infer the \action behavior of all 766 collector peers:

\begin{table}[H]
\small
  \centering
  \begin{tabular}{|l||r|r|r|r|}
\multicolumn{1}{l}{assigned roles:} & \multicolumn{4}{c}{classification result:}\\\hline
alltc:		& \tagger & \silent & \undecided &   none\\\hline
\tagger		&    766 &      0 &         0 &      0\\
\tagger (hidden) &      0 &      0 &	    0 & 72,185\\\hline
  \end{tabular}
\end{table}

The remaining $\sim$72k ASes are hidden behind the those peers and thus fall into \textit{none}.
Next, we look at the \handle behavior in the same scenario:

\begin{table}[H]
\small
  \centering
  \begin{tabular}{|l||r|r|r|r|}
\multicolumn{1}{l}{assigned roles:} & \multicolumn{4}{c}{classification result:}\\\hline
alltc: 	   & \ignore & \cleaner & \undecided &   none\\\hline
\cleaner	  &       0 &      578 &          0 &    124\\
\cleaner (hidden) &       0 &        0 &          0 & 11,829\\
\cleaner (leaf)	  &       0 &        0 &          0 & 60,420\\\hline
  \end{tabular}
\end{table}

From the 766 \tagger collector peers, 578 are followed by other collector peers
in the AS-level graph. Recall, the algorithm first determines the \action behavior
of all peers, and then uses them as \downstream \tagger{s} to determine their \handle behavior.
124 peer ASes (as well as 11,829 transit ASes) are never followed by another peer AS, thus they fall into \textit{none}.
The remaining 64 peer ASes appear as leaf and are included in the set of 60,420
leaf ASes.

\textbf{random}: Next, we look at the \action behavior in the random scenario
where, due to lack of ground truth, we assign roles uniformly at random.
Since there are 72,951 ASes in the data set, each \action role (\tagger, \silent)
occurs approximately $\sim$36k times:

\begin{table}[H]
\small
  \centering
  \begin{tabular}{|l||r|r|r|r|}
\multicolumn{1}{l}{assigned roles:} & \multicolumn{4}{c}{classification result:}\\\hline
random:		& \tagger & \silent & \undecided &   none\\\hline
\tagger		& 22,149 &      0 &         0 &  1,541\\
\silent		&      0 & 21,966 &	    0 &  1,571\\
\multicolumn{5}{l}{\footnotesize\textit{skipped hidden}}
  \end{tabular}
\end{table}

Given the distribution of roles in this scenario, around $\sim$13k \tagger ASes and
$\sim$13k \silent ASes are hidden behind a \cleaner. From the remaining $\sim$23k ASes,
more than 90\% ($\sim$22k) are correctly assigned to the respective
class, while for $\sim$1.5k ASes the algorithm fails to generate pre-existing
knowledge.

\textbf{random+noise}: Thus far, our algorithm is able to correctly infer the \action and \handle behavior
of ASes and avoids (mis)classifying hidden ASes.
In the following, we show how the inferences change when the same random scenario (same seed to randomly assign roles)
is augmented with noise:

\begin{table}[H]
\small
  \centering
  \begin{tabular}{|l||r|r|r|r|}
\multicolumn{1}{l}{assigned roles:} & \multicolumn{4}{c}{classification result:}\\\hline
random+noise:	& \tagger & \silent & \undecided &   none\\\hline
\tagger		& 21,625 &      0 &        1 &  2,064\\
\silent		&     53 &  3,679 &    17,687 &  2,118\\
\tagger (hidden) &      0 &     12 &	    2 & 12,766\\
\silent (hidden) &      1 &     9 &	    3 & 12,931\\\hline
  \end{tabular}
\end{table}

Most notably, $>$80\% (17,687) of the \silent ASes are affected by the addition
noise such that they are labeled \undecided. On the other hand, \tagger ASes
are minimally affected compared to the random scenario (524 correct inferences less).
Also, we observe 53 misclassifications and classifications where the behavior
is hidden.

Similar to \silent ASes in the case of \action behavior, \cleaner ASes are also
affected by noise:

\begin{table}[H]
\small
  \centering
  \begin{tabular}{|l||r|r|r|r|}
\multicolumn{1}{l}{assigned roles:} & \multicolumn{4}{c}{classification result:}\\\hline
random+noise:	& \ignore & \cleaner & \undecided &   none\\\hline
\ignore		   &   2,294 &        0 &         63 &  1,134\\
\cleaner	   &       1 &      738 &      1,647 &  1,153\\
\multicolumn{5}{l}{\footnotesize\textit{skipped hidden and leaf}}
  \end{tabular}
\end{table}

Here, 1,647 (or $\sim$68\%) of \cleaner ASes are labeled \undecided compared
to the random scenario.

\textbf{random-p}: Lastly, we consider the random scenario where around 50\%
of the \tagger ASes are randomly chosen to tag selectively, \ie, not on
provider links but on all other links (including collectors). This results
in the following distribution of \action roles: \selective $\sim$18k, \tagger $\sim$18k,
and around 36k \silent ASes.

\begin{table}[H]
\small
  \centering
  \begin{tabular}{|l||r|r|r|r|}
\multicolumn{1}{l}{assigned roles:} & \multicolumn{4}{c}{classification result:}\\\hline
random-p:	    & \tagger & \silent & \undecided &   none\\\hline
\tagger		    &  6,750 &      0 &         0 &  5,876\\
\silent		    &      0 & 13,445 &         0 & 11,983\\
\selective	    &  2,351 &  3,538 &       837 &  5,984\\
\multicolumn{5}{l}{\footnotesize\textit{skipped hidden}}
  \end{tabular}
\end{table}

In general, the number of not inferred ASes (\textit{none}) increases in all cases compared
to the random scenario. Since \selective ASes are not tagging towards their providers,
fewer communities make it to the collector.
Regarding the visible \tagger and \silent ASes, there are no misclassifications.
However, almost 50\% of the ASes in both cases are not inferred.
Interestingly, from the $\sim$12k visible \selective ASes, around 5.9k are
classified either \tagger or \silent. Only 837 ASes are \undecided.

Summary: The algorithm classifies consistent behavior with a high precision.
It avoids misclassifications of \silent in the presence of noise (\undecided instead),
while \tagger ASes are mostly unaffected. Furthermore, the algorithm avoids
classifying ASes that are hidden or leaf ASes.

%% file: sections/analysis.tex
\section{Analysis}
\label{sec:analysis}

In the previous section, we have introduced an inference method that allows us
to infer the \action and \handle behaviors of ASes. We devised an algorithm
that utilizes routing information, specifically AS paths and community attributes,
and validated its properties using customized data sets.

Next, we show the classification results when the inference method is applied
on actual, unmodified routing data, characterize the involved ASes, and validate our inferences
by actively injecting route announcements with attached communities to the
real-world routing system.

\subsection{Classification results}
\label{sec:analysis-class-results}

\begin{table}[t]
\small
  \centering
  \begin{tabular}{|l|r|r|r|r|r|}
\hline
Input data & \ripe & \rviews 
& \isolario & \datamain & \pch \\\hline\hline
\action: & & & & &\\
\tagger & 723 & 717 & 809 & 860 & 521\\
\silent & 10,594 & 9,902 & 9,901 & 12,315 & 2,214\\
\undecided & 778 & 827 & 902 & 994 & 286\\
\textit{none} & 60,642 & 61,391 & 61,010 & 58,782 & 64,520 \\\hline
\handle: & & & & &\\
\ignore & 201 & 198 & 216 & 271 & 173\\
\cleaner & 285 & 309 & 239 & 417 & 133\\
\undecided & 216 & 137 & 146 & 308 & 85\\
\textit{none} & 72,035 & 72,193 & 72,024 & 71,995 & 67,150\\\hline
\tagger-\ignore & 51 & 70 & 79 & 84 & 68 \\
\tagger-\cleaner & 58 & 64 & 47 & 81 & 37 \\
\silent-\ignore & 88 & 66 & 83 & 107 & 72 \\
\silent-\cleaner & 167 & 173 & 129 & 251 & 70 \\\hline
  \end{tabular}
  \caption{Classification results using real BGP data (RIBs + updates), May 19, 2021. \ripe, \rviews, and \isolario are comparable. Aggregated data set \datamain yields most classifications. \pch includes only updates.}
  \label{tab:classresults-realdata}
\end{table}

Table~\ref{tab:classresults-realdata} summarizes the community usage classification
per route collector project and of the aggregation of \ripe, \rviews, and
\isolario (\datamain). We also include the inferences based on \pch update data
only. Note that the \pch data set does not include RIB information,
see Section~\ref{sec:datasets}. We split the statistics by \action and \handle.
The algorithm exhibits a similar share of identified behavior
among the three collector projects \ripe, \rviews, and \isolario: 717 to 809
ASes exhibit \tagger behavior, 9,901 to 10,594 \silent, 198 to 216 \ignore,
and 239 to 309 ASes exhibit \cleaner behavior.
Looking at the aggregate (\datamain) the numbers are comparable (860, 12,315, 271,
and 417, respectively). \pch shows the least amount of inferred classes.

Regarding the full classification, \ie, both, the \action and the 
\handle for an AS are inferred, we see that \xti (\tagger-\ignore), \xtc 
(\tagger-\cleaner), and \xsi (\silent-\ignore) show similar numbers; between 
47 and 88. \xsc (\silent-\cleaner) is the most common with 129 to 173 
inferred ASes. The aggregated data set shows an increase compared to the 
individual collectors with a total of 523 fully classified ASes out of which 
273 are collector peers. Next, we take a closer look at the stability and 
longitudinal changes of our inferences.

\subsubsection{Stability and longitudinal view}

\begin{figure}[t]
\centering
\includegraphics[width=.49\linewidth]{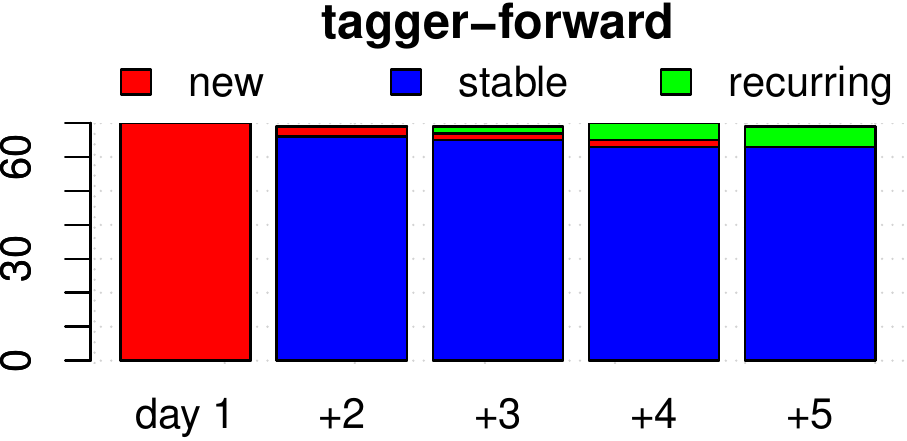}
\includegraphics[width=.49\linewidth]{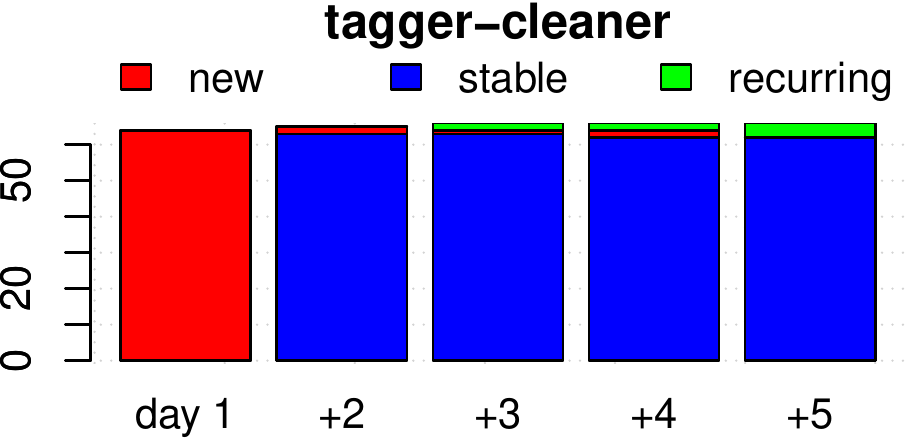}
\includegraphics[width=.49\linewidth]{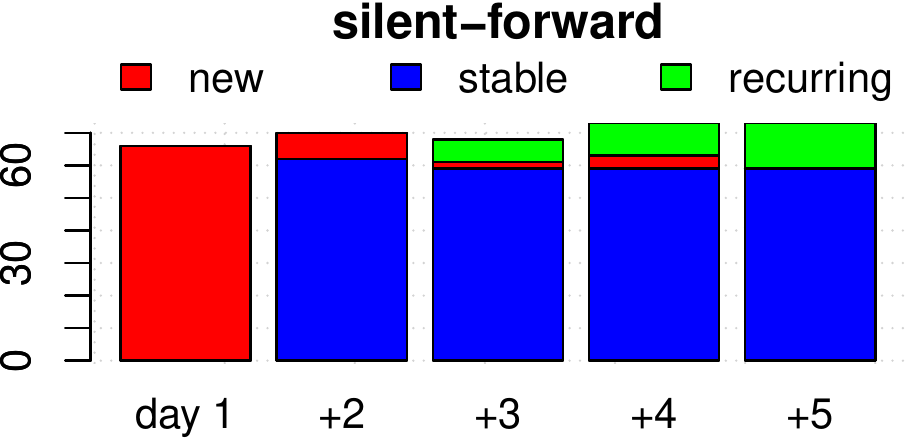}
\includegraphics[width=.49\linewidth]{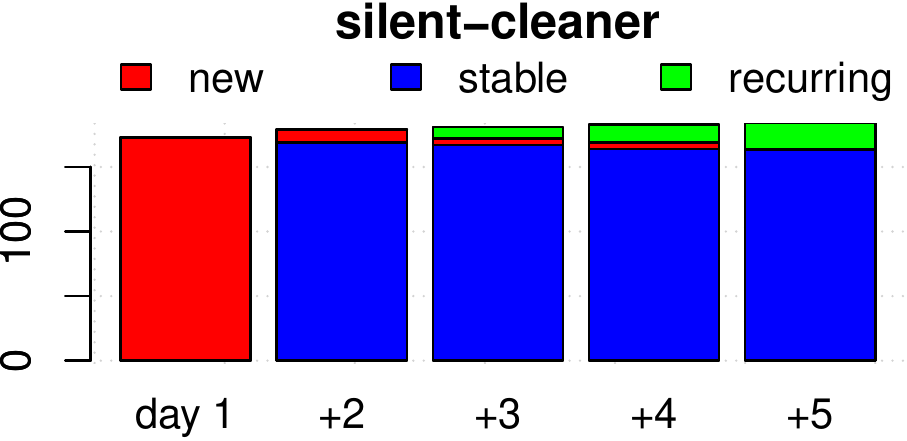}
\caption{Impact of incrementally adding more days to the input data. For each full classification (\xti, \xtc, \xsi, \xsc), barplot shows the total number of ASes (y-axis) and the number of successive days used (x-axis).
\textit{New} ASes appear for the first time in the respective full class, \textit{stable} ASes appear consistently since day 1 (May 19, 2021), and \textit{recurrent}
ASes occur with interruptions. \rviews data only.}
\label{fig:analysis-incremental}
\vspace{-4mm}
\end{figure}

In order to better understand how adding more data impacts our inferences,
we next apply our algorithm to a day worth of data which we incrementally
increase by a successive day for a total of 5 days, starting in May 19, 2021.
Thus, the final data set from May 23 contains data from all 5 days.
Since the inference numbers in Table~\ref{tab:classresults-realdata} for the collectors
and the aggregated \datamain are comparable, in the following we focus only on \rviews
data sets.

Figure~\ref{fig:analysis-incremental} shows the number of \textit{new}, \textit{stable},
and \textit{recurrent} ASes for each full class (\xti, \xtc, \xsi, \xsc). We can
see that except for day 1, only a few ASes are new, \ie, a maximum of 10 at day +2 in the case of \xsc.
Moreover, while we observe some amount of recurring ASes, the majority (90-97\%) of
ASes appear stable since day 1.
We are confident that using our algorithm on a day's worth of routing data is of
sufficient granularity to provide stable inferences.

\begin{figure}[t]
\centering
\includegraphics[width=.99\linewidth]{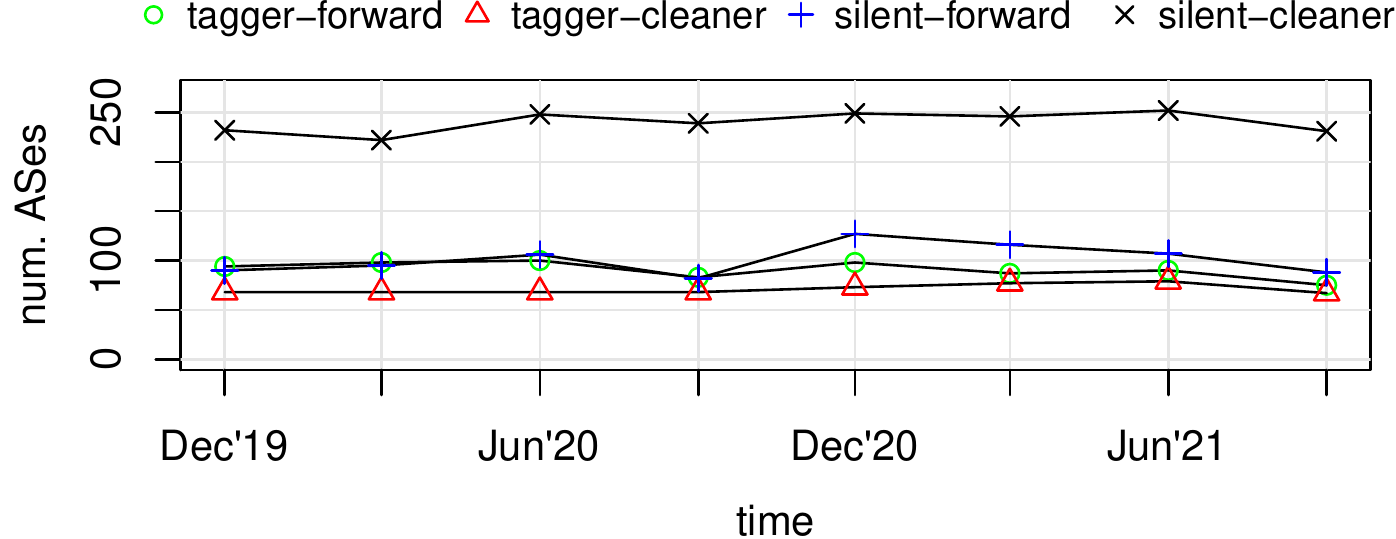}
\vspace{-.5cm}
\caption{Longitudinal view on community usage. For each full classification (\xti,\xtc,\xsi,\xsc), number of total ASes (y-axis) per day plotted over a time period of 2 years,
from December 15, 2019 to (and including) September 15, 2021, every three months. Aggregated data from \ripe, \rviews, and \isolario used.}
\label{fig:analysis-historical}
\vspace{-4mm}
\end{figure}

Next, we investigate how the number of fully classified ASes (\xti, \xtc, \xsi, and \xsc) change over time.
Therefore we use aggregated data sets, \ie, \ripe, \rviews, and \isolario combined,
one day every three months over a time period of two years, see Figure~\ref{fig:analysis-historical}.
The time period ranges from December 15, 2019 to (and including) September 15,
2021. There is no significant increase or decrease in the number of fully classified ASes
discernible. Rather, the number of individual ASes per class and per day are comparable to those
presented in Table~\ref{tab:classresults-realdata} for \datamain.
Overall, these results are indicative of a stable, consistent community usage
behavior throughout the past two years, involving a small set of ASes.

\subsection{Peer ASes}
\label{sec:analysis-peerASes}

\begin{figure}[t]
\centering
\includegraphics[width=.99\linewidth]{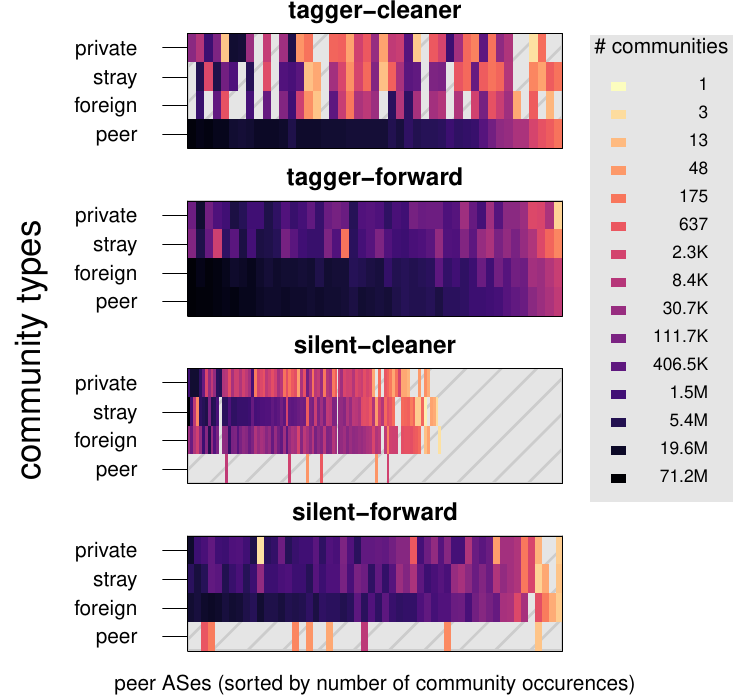}
\caption{Counting different community types (y-axis) at peer ASes (x-axis) with full classification, \ie, \xti, \xtc, \xsi, and \xsc. The ASes are ordered by number of communities and the color scale is logarithmic.}
\label{fig:analysis-peerASes}
\end{figure}

In the following, we take a closer look at 273 collector peers of the 523 
fully classified ASes, see Section~\ref{sec:analysis-class-results}. Since 
peer ASes are connected directly to BGP collectors, their community set 
output ($\coutput(A_1)$) should be not impaired by other ASes along the 
path. Specifically, we investigate whether $\coutput(A_1)$ corresponds to 
the inferred \action and \handle behavior of a peer AS $A_1$. For that, we 
count the type of communities that are included in $\coutput(A_1)$, \ie, 
\textit{peer}, \textit{foreign}, \textit{stray}, and \textit{private}, see 
Section~\ref{sec:background-communities}.

Based on our mental model in Section~\ref{sec:background-commusage}, the 
intuition of our approach is as follows: \textit{peer} communities should 
only be observable when the peer AS is a \tagger. We should not be able to 
observe \textit{peer} communities when the AS is \silent. On the other 
hand, \textit{foreign} should only be visible if the peer AS is \ignore, and 
not visible if that AS is a \cleaner. Note, since our inference method 
discards \textit{stray} and \textit{private} communities 
(Section~\ref{sec:inference-implications}), we expect to see them independent 
of the peer AS classification.

Figure~\ref{fig:analysis-peerASes} shows four scatter plots, one for each
(full) class \xti (\tagger-\ignore), \xtc (\tagger-\cleaner), \xsi (\silent-\ignore),
and \xsc (\silent-\cleaner). In each plot, the x-axis contains
peer ASes of the corresponding class, ordered by the number of total communities,
and the y-axis indicates the community types. The color palette indicates
the number of total communities in a logarithmic scale, ranging from 1 to 71M+.

In accordance with our expectations, we observe a large number \textit{peer}
communities in the cases \xti and \xtc,
indicated by the bottom bar of the respective plots. We observe few to no
\textit{peer} communities in the cases \xsi and \xsc.
Furthermore, we observe a large amount of \textit{foreign} communities in 
the cases \xti and \xsi, \ie, when the peer ASes do not remove communities 
from other ASes in the AS path (see second bottom bar). Conversely, we see 
only few to no \textit{foreign} communities when the peer AS is \xtc or 
\xsc, \ie, the AS is removing communities from other ASes in the AS path. 
However, there are also some \cleaner ASes that, given the number of 
\textit{foreign} communities, contradict their inference. Those 
\textit{foreign} communities can be caused, \eg, by one or multiple 
\downstream ASes that our algorithm was not able to identify as \tagger (and 
thus invalidate \condB).
Last, since we do not 
consider \textit{stray} and \textit{private} communities in our inference 
method, we observe them in all classes, in particular when the peer AS is 
\ignore, but also when it is a \cleaner.

Conclusion: Overall, our observations of community usage at collector peers align with our
mental model and corroborate the correct functioning of our inference method.

\subsection{AS characterization}

Next, we explore the characteristics of ASes based on their
classification. For that we make use of CAIDA's customer cone data sets \cite{caida:as2rel}.
The customer cone of an AS includes itself and all ASes that can be
reached by only traversing customer links. Leaf ASes thus have a customer cone
size of 1. We use the customer cone size as an indicator for the size of an AS.

\begin{figure}[t]
\begin{center}
\includegraphics[width=.91\linewidth]{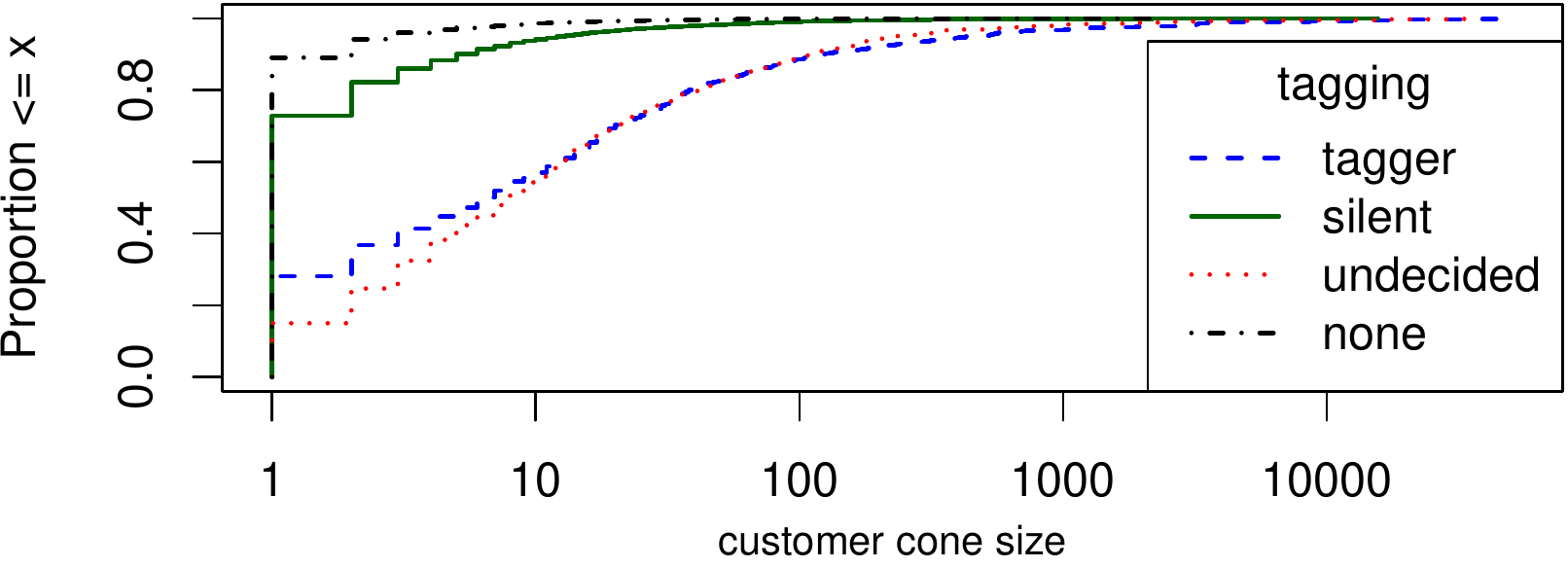}\\
\vspace{.2cm}
\includegraphics[width=.91\linewidth]{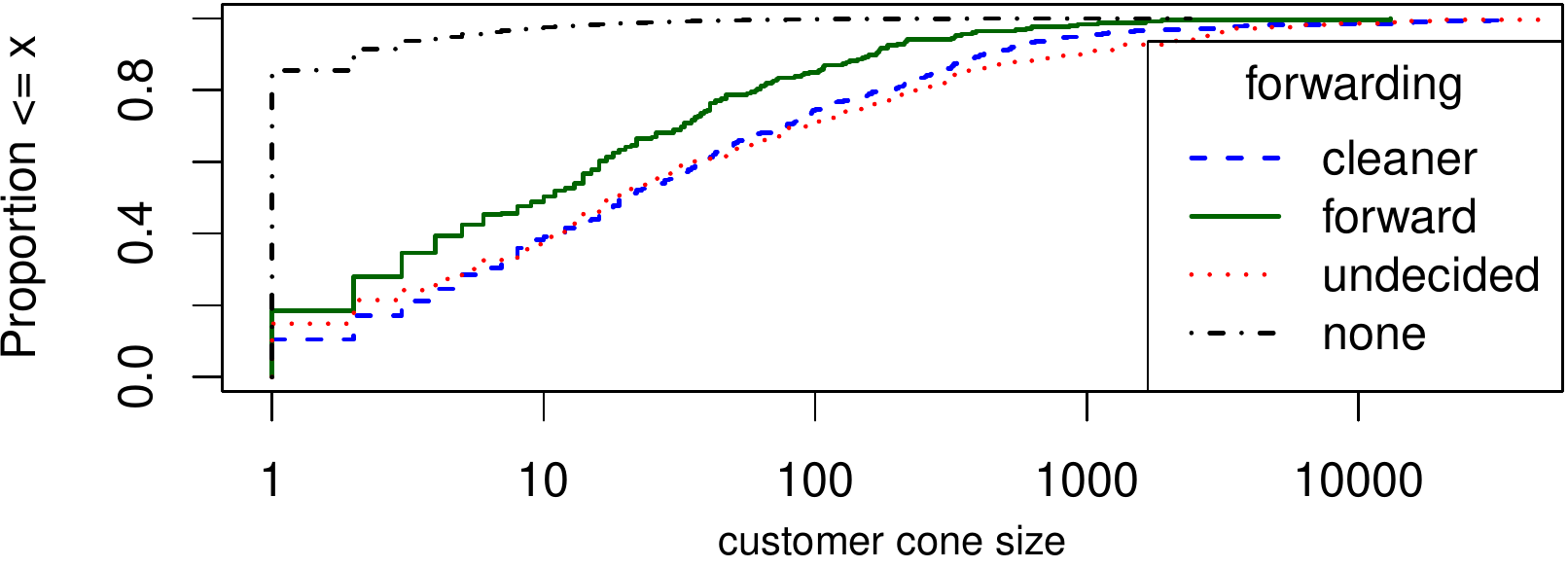}
\vspace{-3mm}
\caption{CDFs showing customer cone size distribution of ASes per \action behavior (top) and \handle behavior (bottom).
All behaviors (except for \silent, and those that are \textit{none}) are attributed to mostly larger and non-leaf ASes.}
\label{fig:analysis-cc-action}
\end{center}
\end{figure}

In Figure~\ref{fig:analysis-cc-action}, we plot CDFs over customer cone sizes
for all ASes in \datamain. Starting with the \action behavior (top plot),
we see a significant difference of cone sizes between
ASes that are classified as \tagger and those that are classified as \silent.
$\sim$70\% of \silent ASes have a cone size of 1, \ie, they are leaf ASes. Also,
while around 10\% of \silent ASes have a cone size greater than 10, it is around
50\% for \tagger ASes, indicating that ASes at the edge of the AS-level graph
usually do not add their own communities. Interestingly, ASes with \undecided
behavior exhibit a similar distribution of cone sizes as \tagger{s}.
Lastly,
ASes with no inference are mostly leaf ASes, \ie in $\sim$90\% of the cases.
Most notably for \handle behavior (bottom plot), \cleaner and \silent ASes show a
similar distribution of cone sizes, indicating that \cleaner and \silent ASes
are common across all AS sizes. Again, ASes with no inference are mostly leaf
ASes.

\subsection{Validation of inferences}

\begin{table}
  \centering
\small
\begin{tabular}{|c|c|c|}\hline
experiment date & communities present & communities not present \\\hline
2021-05-19 & 6/177 (3\%) & 285/367 (78\%) \\
2021-07-15 & 1/104 (1\%) & 286/365 (78\%) \\
2021-08-15 & 0/61 (0\%) & 300/359 (84\%) \\\hline
\end{tabular}
\caption{PEERING experiments: Table shows the share of paths containing
at least one \cleaner AS, when a) communities are present and b) when
communities are not present. We perform three temporally uncorrelated experiments.}
\label{tab:peering}
\vspace{-4mm}
\end{table}

Finally, we validate our inferences with external ground truth information obtained by
injecting route announcements with consistent use of communities. Thus, by having
control over a \tagger, we check how often the observation of our communities
contradicts the inferences of AS behaviors along the corresponding AS path.
For example, given an AS path and a community set that includes our communities,
the AS path must not include a \cleaner AS, otherwise the community would have
been not visible (see Section~\ref{sec:inference-implications}).
Similarly, if the community set does not include any of our
communities, there must be at least one \cleaner in the AS path.

We utilized the PEERING testbed to conduct our real-world
validation~\cite{schlinker2019peering}.  Using PEERING, we 
announce a /24 prefix via 12 available Points of Presence (PoPs) 
for approximately one week beginning on May 19, 2021.
To each PoP, we add a unique pair of communities to the announcement;
these communities use PEERING's 47065 ASN in the upper 2 bytes and
unique values in the lower 2 bytes.  Via these PoPs, PEERING had
460 active BGP peers in total at the time of our experiment.

We extract the relevant announcements from our data set \datamain by filtering
for the prefix. We observe a total of 7,503 announcements corresponding to that
prefix. From the 12 community sets we have configured, only 6
appear in around 30\% of those announcements.
From the observed announcements, we extract 549 unique AS paths.
Interestingly, only 5 AS paths are not consistently
attached with our communities, \ie, they appear with and without our communities.
The remaining 544 AS paths are either always, or never, associated
with our communities.
This indicates a consistent usage of communities along the same path.

In order to validate our inferences, we count how often a \cleaner exists
in the AS paths, when a) the corresponding community set contains our communities,
and when b) the community set \textit{does not} contain our communities.
We first look at 367 unique tuples that do not include any of our communities:
in 285 cases ($\sim$78\%) at least one \cleaner exists in the AS path.
81 cases ($\sim$22\%) include at least one AS that exhibits \undecided behavior.
Only 1 AS path does not include an identified \cleaner and thus contradicts our
inferences.
Next, we select 177 unique AS path / community set tuples that \textit{do} contain our communities:
Here, 6 (3\%) AS paths contradict our inferences, \ie, the AS path contains a
\cleaner AS. However, we note that 152 AS paths ($\sim$85\%) include at least
one AS with \undecided behavior. The remaining 19 AS paths include neither.
In Table~\ref{tab:peering} we show a summary of this and two
additional experiments we have performed, with similar outcomes.

\vspace{-.5cm}

%% file: sections/conclusion.tex
\section{Conclusions and Future Work}
\label{sec:conclusion}

In this work, we take a step toward achieving a more grounded
understanding of BGP community usage behavior at a per-AS
granularity.  While our algorithm is inferential and its
ability to classify an AS's community usage depends both
upon BGP visibility and the complexity of the AS's policies,
ours is the first work to take on this challenging task.  Furthermore,
through extensive testing and validation, we show that our 
algorithm has high precision -- when it does make an inference,
it is generally correct.

We apply our algorithm to large-scale, real-world BGP data from the
major route collectors in order to maximize coverage and make
inferences for the most ASes possible.  We characterize the 
ASes exhibiting different usage behaviors and find that 
\tagger, \ignore and \cleaner ASes typically have a large customer
cone, while \silent are typically at the edge.

We publicly release our algorithm and inferences to the community
to support related BGP and network research, including automated
community processing in networks based on inferences, Internet 
modeling, and improved BGP security posture~\cite{url:communityusage}.

In future work, we plan to extend and improve the algorithm,
especially removing the strict assumption that community
tags use the ASN of the network that added the community.  In
this fashion, we wish to identify not only whether an AS is
a \tagger, but also \emph{which} communities it adds.  This
ability will be especially useful to differentiate signaling
versus informational communities, and to support efforts directed
toward automated and safe community filtering.

%% file: sections/acknowledgements.tex
\subsection*{Acknowledgements}

We would like to thank our shepherd, Neil Spring, and the anonymous reviewers
for useful and constructive feedback.
We would also like to thank Italo Cunha and Ethan Katz-Bassett for providing 
access to the PEERING test bed.
This work has been partially funded by NSF grant CNS-1855614,
the European Research Council (ERC) Starting Grant ResolutioNet
(ERC-StG-679158), BMBF BIFOLD 01IS18025A and 01IS18037A,
and performed while the first author held an NRC
Research Associateship award at the Naval Postgraduate School.
Views and conclusions are those of the authors and should not be
interpreted as representing the official policies or position of the
U.S. government, the NSF, ERC, BMBF, or NRC.

%% file: sections/appendix.tex
\appendix

\section{Appendix}
\label{sec:appendix}

\begin{lstlisting}[language=Python, label={lst:algorithm}, caption={Inference algorithm, iterating over \pcpair pairs by path index $x$, where $path = A_1, A_2, ..., A_n$, and $comm = output(A_1)$.}, float=h, floatplacement=f, mathescape=true, escapechar=\%]
for x=1,...,N:
  # PHASE 1: count %\action%
  for each %\pcpair%:
    assert(cond1)
    t[$A_x$]++ if $A_x$:* in comm
    s[$A_x$]++ if $A_x$:* not in comm
  # PHASE 2: count %\handle%
  for each %\pcpair%:
    assert(cond1 and cond2)
    i[$A_x$]++ if $A_t$:* in comm
    c[$A_x$]++ if $A_t$:* not in comm
\end{lstlisting}

\begin{lstlisting}[language=Python, label={lst:algorithm-alter}, caption={Alternative inference algorithm, iterating over \pcpair pairs without conditions, where $path = A_1, A_2, ..., A_n$, and $comm = output(A_1)$.}, float=h, floatplacement=t, mathescape=true, escapechar=\%]
# INIT
aslist[] = %\silent%|%\ignore%

# PHASE 1: count %\action%
for each %\pcpair%:
  for each i = 1,...,n:
    t[$A_x$]++ if $A_x$:* in comm
    s[$A_x$]++ if $A_x$:* not in comm
# PHASE 2: count %\handle%
for each %\pcpair%:
  for each x = n-1,...,1:
    c[$A_x$]++ if $A_{x+1}$ not in comm
    else:
      i[$A_j$]++ for j in 1,...,$A_{x+1}$
\end{lstlisting}

\begin{table}[H]
\small
  \centering
  \begin{tabular}{|l||r|r|r|r|}
\multicolumn{1}{l}{assigned roles:} & \multicolumn{4}{c}{classification result:}\\\hline
alltf:		& \tagger & \silent & \undecided &   none\\\hline
\tagger		& 69,997 &      0 &         0 &  2,954\\\hline
\multicolumn{5}{l}{}\\\hline
alltc:		& \tagger & \silent & \undecided &   none\\\hline
\tagger		&    766 &      0 &         0 &      0\\
\tagger (hidden) &      0 &      0 &	    0 & 72,185\\\hline
\multicolumn{5}{l}{}\\\hline
random:		& \tagger & \silent & \undecided &   none\\\hline
\tagger		& 22,149 &      0 &         0 &  1,541\\
\silent		&      0 & 21,966 &	    0 &  1,571\\
\tagger (hidden) &      0 &      0 &	    0 & 12,780\\
\silent (hidden) &      0 &      0 &	    0 & 12,944\\\hline
\multicolumn{5}{l}{}\\\hline
random+noise:	& \tagger & \silent & \undecided &   none\\\hline
\tagger		& 21,625 &      0 &        1 &  2,064\\
\silent		&     53 &  3,679 &    17,687 &  2,118\\
\tagger (hidden) &      0 &     12 &	    2 & 12,766\\
\silent (hidden) &      1 &     9 &	    3 & 12,931\\\hline
\multicolumn{5}{l}{}\\\hline
random-p:	    & \tagger & \silent & \undecided &   none\\\hline
\tagger		    &  6,750 &      0 &         0 &  5,876\\
\silent		    &      0 & 13,445 &         0 & 11,983\\
\selective	    &  2,351 &  3,538 &       837 &  5,984\\
\tagger (hidden)    &      0 &      0 &         0 &  5,562\\
\silent (hidden)    &      0 &      0 &	        0 & 11,082\\
\selective (hidden) &      0 &      0 &         0 &  5,543\\\hline
\multicolumn{5}{l}{}\\\hline
random-pp:	   & \tagger & \silent & \undecided &   none\\\hline
\tagger		   &  2,163 &      0 &         0 &  10,463\\
\silent		   &      0 &  4,429 &         0 & 20,999\\
\selective	   &  1,301 &    713 &       134 &   10,562\\
\tagger (hidden)	   &      0 &      0 &         0 &  5,562\\
\silent (hidden)    &      0 &      0 &	       0 & 11,082\\
\selective (hidden) &      0 &  0 &         0 & 5,543\\\hline
  \end{tabular}
  \caption{Assigned roles vs. classification results: Confusion Matrices for \action behavior per scenario.}
  \label{tab:confusion-action}
\end{table}

\begin{table}[H]
\small
  \centering
  \begin{tabular}{|l||r|r|r|r|}
\multicolumn{1}{l}{assigned roles:} & \multicolumn{4}{c}{classification result:}\\\hline
alltf: 	   & \ignore & \cleaner & \undecided &   none\\\hline
\ignore		   &  10,427 &        0 &          0 &  2,104\\
\ignore (leaf)	   &       0 &        0 &          0 & 60,420\\\hline
\multicolumn{5}{l}{}\\\hline
alltc: 	   & \ignore & \cleaner & \undecided &   none\\\hline
\cleaner	  &       0 &      578 &          0 &    124\\
\cleaner (hidden) &       0 &        0 &          0 & 11,829\\
\cleaner (leaf)	  &       0 &        0 &          0 & 60,420\\\hline
random: 	   & \ignore & \cleaner & \undecided &   none\\\hline
\ignore		   &   2,400 &        0 &          0 &  1,091\\
\cleaner	   &       0 &    2,433 &          0 &  1,106\\
\ignore (hidden)   &       0 &        0 &          0 &  2,750\\
\cleaner (hidden)  &       0 &        0 &          0 &  2,750\\
\ignore (leaf)	   &       0 &        0 &          0 & 30,335\\
\cleaner (leaf)    &       0 &        0 &          0 & 30,085\\\hline
\multicolumn{5}{l}{}\\\hline
random+noise: 	   & \ignore & \cleaner & \undecided &   none\\\hline
\ignore		   &   2,294 &        0 &         63 &  1,134\\
\cleaner	   &       1 &      738 &      1,647 &  1,153\\
\ignore (hidden)   &       0 &        2 &          2 &  2,746\\
\cleaner (hidden)  &       0 &        1 &          0 &  2,750\\
\ignore (leaf)	   &       0 &        0 &          0 & 30,335\\
\cleaner (leaf)    &       0 &        0 &          0 & 30,085\\\hline
\multicolumn{5}{l}{}\\\hline
random-p:	   & \ignore & \cleaner & \undecided &   none\\\hline
\ignore		   &     925 &       75 &        266 &  1,666\\
\cleaner	   &       0 &    1,355 &          0 &  1,663\\
\ignore (hidden)   &       0 &       52 &          0 &  3,265\\
\cleaner (hidden)  &       0 &       47 &          0 &  3,217\\
\ignore (leaf)	   &       0 &        0 &          0 & 30,480\\
\cleaner (leaf)    &       0 &        0 &          0 & 29,940\\\hline
\multicolumn{5}{l}{}\\\hline
random-pp:	   & \ignore & \cleaner & \undecided &   none\\\hline
\ignore		   &     221 &       72 &        279 &  2,341\\
\cleaner	   &       0 &      610 &          0 &  2,386\\
\ignore (hidden)   &       0 &       14 &          0 &  3,322\\
\cleaner (hidden)  &       0 &       19 &          0 &  3,267\\
\ignore (leaf)     &       0 &        0 &          0 & 30,480\\
\cleaner (leaf)    &       0 &        0 &          0 & 29,940\\\hline
  \end{tabular}
  \caption{Assigned roles vs. classification results: Confusion Matrices for \handle behavior per scenario.}
  \label{tab:confusion-handle}
\end{table}